\documentclass[12pt]{article} 
\pdfoutput=1
\newlength{\abstractwidth}
\usepackage{color}
\usepackage{graphicx}

\renewcommand{\thefootnote}{\fnsymbol{footnote}}
\renewcommand{\thanks}[1]{\footnote{#1}}
\newcommand{\starttext}{
\setcounter{footnote}{0}
\renewcommand{\thefootnote}{\arabic{footnote}}}

\newcommand{\bea}{\begin{eqnarray}}
\newcommand{\eea}{\end{eqnarray}}
\newcommand{\ee}{\end{equation}}
\newcommand{\be}{\begin{equation}}

\newcommand{\sm}{\smallskip}

\def\cO{{\cal O}}

\def\p{\partial}

\def\l({\left(}
\def\r){\right)}

\def\a{\alpha}

\def\ep{\varepsilon}

\def\eps{\epsilon}

\def\l{\lambda}

\def\f{\varphi}
\def\G{\Gamma}

\def\sh{{\rm sh}}

\def\Bh{\hat{B}}
\def\Th{\hat{T}}
\def\sh{\hat{s}}
\def\rt{\rightarrow}

\def\no{\nonumber}

\begin{document}
\starttext
\setcounter{footnote}{0}

\begin{flushright}
\today
\end{flushright}

\bigskip

\begin{center}

{{\Large \bf  Magnetic Field Induced Quantum Criticality \\ via new 
Asymptotically AdS$_5$ Solutions} 
\footnote{This work was supported in part by NSF grant PHY-07-57702.}}

\vskip .4in

{\large \bf Eric D'Hoker and  Per Kraus}

\vskip .2in

{ \sl Department of Physics and Astronomy }\\
{\sl University of California, Los Angeles, CA 90095, USA}\\
{\tt \small dhoker@physics.ucla.edu; pkraus@ucla.edu}

\end{center}

\vskip .2in

\begin{abstract}

\vskip 0.1in 

Using analytical methods, we derive and extend previously obtained numerical results on
the low temperature properties of holographic duals to four-dimensional gauge theories at
finite density in a nonzero magnetic field.   We find a new asymptotically AdS$_5$ solution
representing the system at zero temperature.   This solution has vanishing entropy density, 
and the charge density in the bulk is carried entirely by fluxes.  The dimensionless magnetic 
field to charge density ratio for these solutions is bounded from below, with a quantum critical 
point appearing  at the lower bound. Using matched asymptotic expansions, we extract the 
low temperature thermodynamics of the system.  Above the critical magnetic field, the low 
temperature entropy density takes a simple form, linear in the temperature, and with a specific heat 
coefficient diverging at the critical point.  At the critical magnetic field, we derive the scaling 
law $s \sim T^{1/3}$  inferred previously from numerical analysis. 
We also compute the full scaling function describing the region near the critical point, and identify 
the dynamical critical exponent: $z=3$.

These solutions are expected to holographically represent boundary theories in which strongly 
interacting fermions are filling up a Fermi sea. They are fully top-down constructions in which
both the bulk and boundary theories have well known embeddings in string theory.

\end{abstract}

\newpage


\newpage

\section{Introduction and summary of results}
\setcounter{equation}{0}

Understanding the properties of interacting fermions at finite density is a central problem in diverse
areas of physics, ranging from condensed matter  to nuclear and astrophysics. 
The dual gravity description arising in the AdS/CFT correspondence may provide
a new window on the  strongly coupled versions of such systems, and the last few years
have seen significant effort in this direction.  

\sm

The model that has attracted the most attention consists of studying bulk fermions propagating
on a Reissner-Nordstrom black brane background 
\cite{Liu:2009dm,Cubrovic:2009ye,Faulkner:2009wj,Faulkner:2010tq}.  
The advantage of this model is that it
is very simple, yet the fermion correlators display  interesting non-Fermi liquid behavior.  
But there are some serious limitations.   In general terms, this construction does not harness
the full power of gauge-gravity duality, as the fermions are described ``explicitly" in the bulk rather
than holographically. 
In describing boundary fermions we would like gravity to do the full path integral for us (at large $N$), but
in this setup the task of studying quantum fluctuations of the fermions  is left to the user. 
Another problematic feature is that the non-Fermi liquid behavior of the correlators hinges on the
existence of a near horizon AdS$_2$ factor with  finite ground state entropy density.  This makes the
nature of the dual field theory rather obscure, and seemingly quite different from the real world
systems with which one is hoping to make contact.   Other approaches to studying holographic
fermions, with various pros and cons, include 
\cite{Sakai:2004cn,Davis:2008nv,Rey:2008zz,Kulaxizi:2008jx}  

\sm 

In fact, a promising holographic  setup for studying interacting fermions at finite density has been
sitting right under our noses, in the form of the duality between 5-dimensional 
Einstein-Maxwell-Chern-Simons theory and 4-dimensional supersymmetric gauge theories.   
Simply turning on a finite
charge density in this system is inadequate, as it leads to the same finite entropy density ground state 
Reissner-Nordstrom geometry noted above.  This is presumably somehow a reflection of the fact that the
gauge theory contains massless charged bosons, which will want to condense at nonzero chemical
potential, precluding the formation of a Fermi surface.    However, turning on a magnetic field
removes these problems:  in the gauge theory the bosonic modes are lifted up in energy above 
the lighest fermion modes, and in the bulk a smooth zero entropy ground state geometry emerges.
The presence of the magnetic field also provides access to a tunable parameter and, just as in the
real world, it is interesting to see how the physical properties change as a function of the field strength. 

\sm

\begin{figure}[h!]
\begin{centering}  
\includegraphics[scale=.6]{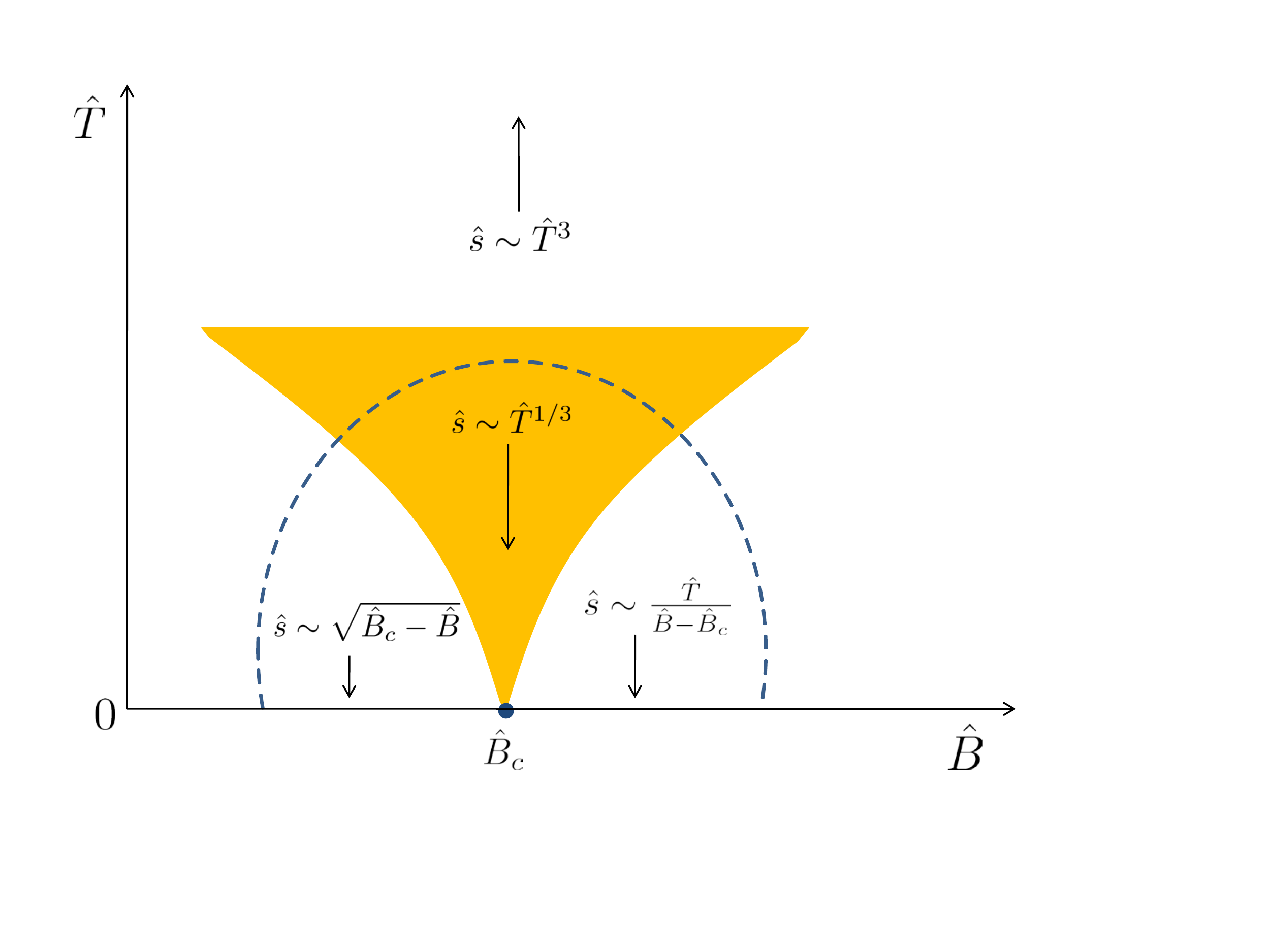}
\caption{Schematic phase diagram illustrating the various  behaviors of the entropy density 
versus temperature and magnetic field. The region inside the dashed line is controlled by the 
quantum critical point at $(\Th=0,\Bh=\Bh_c)$, and  the entropy density  can be 
expressed in terms of a single scaling function $f$ of $(\hat B - \hat B_c)/T^{2/3}$. 
We move around inside this region by changing the temperature $\Th$ and the relevant coupling 
$\Bh-\Bh_c$.   The boundary of the region is defined to be where  irrelevant operators
become important.  The yellow region denotes a regime where temperature is the largest 
energy scale, corresponding to  the argument of the scaling function $f$ being small.  Outside
the yellow region the low temperature behavior of the entropy density, for fixed $\hat B$, 
is either constant or linear in $\hat T$, 
depending on whether the quantum critical point is approached from below or 
from above $\hat B_c$ as $\Th\rightarrow 0$.  }
\label{phase}
\end{centering}
\end{figure}

In previous papers \cite{D'Hoker:2009mm,D'Hoker:2009bc,D'Hoker:2010rz} we have initiated a 
detailed study of this system, and a  rich structure
has emerged.   Perhaps most interestingly,  precision numerical analysis revealed the existence 
of a quantum critical point \cite{D'Hoker:2010rz}.  The results are summarized in Fig. \ref{phase} (taken from    \cite{D'Hoker:2010rz}). 
The system is studied as a function
of the dimensionless magnetic field to charge density ratio: $\Bh = B/\rho^{2/3}$.  The thermodynamics
properties are expressed in terms of the dimensionless temperature $\Th$ and entropy density
$\sh$.   For sufficiently large magnetic field, we observed in \cite{D'Hoker:2010rz} a linear 
dependence of the  low temperature entropy density, $\sh = c(\Bh) \Th$.   As the magnetic
field was decreased, or equivalently the charge  density increased, the coefficient $c(\Bh)$ was
found to diverge at a critical value $\Bh= \Bh_c$.    At this critical point a new scaling law for the 
entropy emerged, $\sh \sim \Th^{1/3}$.  Departures from the critical point with respect
to both temperature and magnetic field could be expressed in terms of a single scaling function, 
as is familiar from the study of both classical and quantum critical phenomena.   The numerical
results suggested that we identify the critical point as having dynamical critical exponent $z=3$,
and possessing a relevant operator of scaling dimension $2$.  

\sm
 
Several considerations render this behavior especially appealing. 
In terms of the connection to physical materials, we note that  tuning a magnetic field is a common 
way of locating quantum critical points in heavy fermion
compounds \cite{LRMW}, and the appearance of the critical point closely parallels what we 
observe.\footnote{As discussed in \cite{D'Hoker:2010rz}  there is also a close connection
with the results described in \cite{RPMMG}.}  Namely,
away from the critical point one has  a linear entropy density versus temperature law due to the presence
of a Landau-Fermi liquid, but the coefficient diverges at the critical point, giving way to a non-linear
relation.    Another helpful aspect is that our behavior is arising in a fully ``top-down" setup: the
gauge theories are explicitly known, and include as one example ${\cal N}=4$ super-Yang-Mills theory. 

\sm

On the other hand, since the analysis in  \cite{D'Hoker:2010rz} was heavily numerical the basic 
mechanism underlying the observed behavior was not entirely clear.  Also, numerical results 
were obtained only for a particular value of the bulk Chern-Simons coupling, namely that dictated
by supersymmetry.    To both understand and extend our previous results an analytical treatment is 
clearly desirable. This is the basis of the present paper.

\subsection{Results} 

We have found a new asymptotically AdS$_5$  solution, which is at zero temperature and corresponds to  
a nonzero magnetic field and charge density in the boundary theory.  In a convenient set of 
coordinates the solution takes the form\footnote{By a coordinate transformation the asymptotic
metric may be put in canonical AdS$_5$ form.}
\bea
ds^2 &=& {dr^2 \over L^2} +M dt^2 + 2L dt dx_3 + e^{2V}(dx_1^2+ dx_2^2) \\ \no
F&=& b dx_1 \wedge dx_2 + E dr \wedge dt
\eea
Here the magnetic field $b$ is a constant while the functions $(L,M,V,E)$ depend on $r$ only.  
The explicit solution can
be written almost entirely analytically: the function $V$ must be determined numerically, but it obeys
an equation with no dependence on free parameters and so is determined once and for all, and then
the remaining functions may be solved by quadrature in terms of $V$.   In the near horizon limit 
the functions take the simple form
\bea\label{fncts}
L = 2br~,\qquad M =-\tilde{\alpha} r -{q^2r^{2k}  \over k(k-{1\over 2})} ~,\qquad e^{2V}
= {b\over \sqrt{3}}~,\qquad E = q r^{k-1}
\eea
where $k$ is the coefficient of the Chern-Simons term in the bulk action, and $\tilde{\alpha}$ is a constant. The $(r,t,x_3)$ part of this near horizon metric can be 
recognized as the three-dimensional part of the Schrodinger spacetime used in non-relativistic
holography \cite{Son:2008ye,Balasubramanian:2008dm}.  It is also a solution of three-dimensional 
gravity with a gravitational Chern-Simons term, and
has been studied in this context recently in \cite{Anninos:2010pm}.

\sm

Viewing this solution  as the  zero temperature limit of a  finite temperature solution, we demand
that $M$ be negative near the horizon.   For $k>1/2$, this implies that
 the parameter $\tilde{\alpha}$ must be
non-negative.   As we show, this inequality translates into a bound on the dimensionless magnetic
field to charge density ratio measured at the asymptotic AdS$_5$ boundary.      
We find  $\Bh \geq \Bh_c$ where $\Bh_c$ is a $k$-dependent number that can be 
computed in terms of integrals involving the
$V$ function.   This explains the appearance of a critical magnetic field in our 
previous numerical work  \cite{D'Hoker:2010rz}, and a quantum critical point at $\hat B = \hat B_c$.     From the form of $M(r)$ in (\ref{fncts}) 
it is evident that $k=1/2$ is special,
and indeed we  find that $\Bh_c$ diverges as $k \rt 1/2$.   The quantum critical point is  absent
for $k \leq 1/2$. 

\sm
 
We extract the low temperature thermodynamics by perturbing around the zero temperature 
solutions.   The appropriate technique employs a matched asymptotic expansion, where we 
match two different perturbations, growing away from the horizon and from the AdS$_5$ region
respectively.   This analysis provides formulas for the low temperature entropy density, and gives us
the dynamical critical exponent since it involves scaling the near horizon time and space coordinates
as the temperature is scaled.  

\sm

Above the critical magnetic field this analysis yields the simple result
\bea
\label{entropy1}
\sh = {\pi \over 6}\left( {\Bh^3 \over \Bh^3 -\Bh_c^3}\right) \Th \hskip 1in \Th \rt 0~,\quad   \Bh > \Bh_c 
\eea 
This result displays all the features that arose in our previous numerical analysis. The linear
behavior in temperature is characteristic of a Landau-Fermi liquid, here presumably arising 
from the fact that in the field theory we are filling up states in the lowest fermionic Landau level.
Recall that the specific heat $C$ obeys the same behavior, since $C = T \p s / \p T$. 
In the large $\Bh$ limit the result  $\sh = \pi \Th/6$ agrees with what we found for the purely 
magnetic solutions in \cite{D'Hoker:2009mm}.  In this limit the near horizon geometry is 
BTZ $\times R^2$, and the result corresponds to using the Brown-Henneaux
central charge \cite{Brown:1986nw}  and the Cardy formula.   Away from this limit, 
this approach is not applicable, as the 
near horizon solutions are no longer BTZ black holes on account of the nonzero charge density. 
As we approach $\Bh_c$ from above, we see that the coefficient of the linear term diverges, as was
seen numerically in \cite{D'Hoker:2010rz}.   This is directly analogous to the observation of a
divergence in the specific heat coefficient in real materials approaching a field tuned quantum critical
point.  A finite temperature version of a magnetic field induced phase transition was studied 
holographically in  \cite{Lippert}, and for zero temperature versions involving probe branes 
see \cite{Evans:2010iy,Jensen:2010vd}.

\sm

Tuning to the quantum critical point corresponds to setting $\tilde{\alpha} =0$ in (\ref{fncts}).  The details of the
matched asymptotic expansion now change, and the corresponding result for the low temperature entropy density is 
\bea
\sh = a \Th^{1/3}~,\quad\quad  a^3  = { \pi \over 576 k\hat B_c ^3}\quad\quad\quad   \Bh= \Bh_c   
\eea
This is in accord with the numerical results obtained in 
\cite{D'Hoker:2010rz}.      The scaling analysis shows that the dynamical critical exponent is now
$z=3$.\footnote{For reasons that will be explained in section 4.3.6 below,
this scaling relation holds for all $k \geq 3/4$. For $1/2<k<3/4$
a different scaling relation holds, to be examined elsewhere \cite{inprog}.}

\sm

We can perturb around the critical point by including both a finite temperature and a deviation
in the magnetic field away from the critical value.  This leads to the scaling form for the
entropy density
\bea\label{scform}
\sh = \Th^{1/3} f\left( {\Bh - \Bh_c\over \Th^{2/3}}\right) 
\eea
where the function $f$ can be found in terms of the solution to a cubic equation,
\bea
 f(x)^3 + { x f(x) \over 32k\hat B_c^4}  = a^3
\eea  
Given the result $z=3$, this shows that $\Bh - \Bh_c$ represents a relevant 
coupling with scale dimension $2$ at the fixed point.  Again, we have agreement 
with the numerics in \cite{D'Hoker:2010rz}.  The behavior of other quantities, such as
the magnetization, near the critical point may be obtained from (\ref{scform}) by applying
standard thermodynamic relations. 

\sm

In other studies of quantum critical points in the AdS/CFT correspondence, nontrivial 
dynamical critical exponents are associated with near horizon Lifschitz metrics \cite{Kachru:2008yh}.  
Our mechanism is different, and is tightly linked with the fact that our solutions are
stationary but not static.   Entropy density and temperature are measured at the horizon, while
time and space in the field theory are measured at the AdS$_5$ boundary.   If the natural time
and space coordinates at the horizon and at the boundary are nontrivially related, as is the case here, 
new dynamical scaling laws can ensue.  This same mechanism can be expected to play a role
in other AdS/CMT applications. 

\sm

A noteworthy feature  of our zero temperature solutions is that the charge density measured at infinity is
carried entirely by fluxes in the bulk; some other recent solutions with this property include \cite{Horowitz:2009ij,Bobev:2010de}.  This is possible by virtue of the Chern-Simons coupling, and 
is in contrast to the Reissner-Nordstrom solution, where all the charge is hidden behind the horizon.
The fermions in the boundary theory are thus described fully holographically as classical bosonic
fields in the bulk, and the charge is also accessible since it is outside the horizon.   It will be interesting
to study the dynamical response of these holographic fermions by studying perturbations at nonzero
frequency and wavelength.

\sm

The remainder of this paper is organized as follows. In section 2, we review the Ansatz, 
reduced field equations, and first integrals, in the presence of a uniform magnetic field
and electric charge density, exhibit the allowed near-horizon solutions, and review the 
physical parameters of the problem. In section 3, we derive our new charged asymptotically 
$AdS_5$ solutions, and derive a formula for the critical magnetic field $\hat B_c$ as a 
function of $k$. In section 4, we develop a procedure for matching the perturbation
theories around the near-horizon and asymptotically $AdS_5$ regions, and use this
to derive the low temperature formula for the specific heat coefficient as a function of 
$\hat B$, and the scaling function $f$  near the quantum critical point. In section 5, we 
present a detailed technical discussion of this matching procedure, and solve the 
associated perturbation theories.
In Appendix A, we derive the most general near-horizon solutions. In Appendix B,
we present a detailed discussion of the solution of the more involved equations for 
the asymptotically $AdS_5$ perturbation theory.

\section{Field equations and simple analytic solutions}
\setcounter{equation}{0}

The action for 5-dimensional Einstein-Maxwell theory, with negative cosmological
constant and Chern-Simons term, is given by\footnote{Conventions:
$R^\lambda_{~\mu\nu\kappa}=
\p_\kappa \Gamma^\lambda_{\mu\nu}
-\p_\nu \Gamma^\lambda_{\mu\kappa}
+\Gamma^\eta_{\mu\nu}\Gamma^\lambda_{\kappa\eta}
-\Gamma^\eta_{\mu\kappa}\Gamma^\lambda_{\nu\eta}$,  $R_{\mu\nu} = R^\lambda_{~\mu \lambda \nu}$ and $R=R^\mu_{~\mu}$.  }
\bea
\label{action}
S_{{\rm EM}} = - { 1 \over 16 \pi G_5} \int d^5x \sqrt{-g}
\left ( R + F^{MN} F_{MN} - {12 \over L^2} \right ) + S_{{\rm CS}} + S_{{\rm bndy}}
\eea
where the Chern-Simons term is  
\bea
\label{CS}
S_{{\rm CS}} = {k\over 12\pi G_5} \int  A\wedge F\wedge F
\eea
For the value $k=2/\sqrt{3}$, the action coincides with the bosonic part of $D=5$
minimal gauged supergravity. In this paper, however,  $k$ will often be kept general,
thus allowing for values different from the supersymmetric case.  
We assume $k\geq 0$ without loss of generality, since a sign reversal of $k$
is equivalent to a parity transformation.  Boundary terms in the action
are required for finiteness of the action and the existence of a well posed variational 
problem \cite{Henningson:1998gx,Balasubramanian:1999re};  their explicit form may 
be found in \cite{D'Hoker:2009bc}.   We henceforth set $L=1$. 

\sm

The Bianchi identity is $dF=0$, while the field equations are given by,
\bea\label{FE}
0 & = &  d * F + k F \wedge F
\no \\
R_{MN} & =  &
4  g_{MN} +{1 \over 3} F^{PQ}F_{PQ} g_{MN} -2 F_{MP}F_N{}^P
\eea
With $k=2/\sqrt{3}$,  the action  (\ref{action}) is a consistent truncation known to describe all
supersymmetric compactifications of Type IIB or M-theory to AdS$_5$
\cite{Buchel:2006gb,Gauntlett:2006ai,Gauntlett:2007ma}.  This means that solutions of 
(\ref{FE}) are guaranteed to be solutions of the full 10 or 11 dimensional field equations 
(although for non-supersymmetric solutions there is no guarantee of stability).  
It also implies that the solutions we find
are holographically dual not just to ${\cal N}=4$ super-Yang-Mills, but to the infinite class of 
supersymmetric field theories dual to these more general supersymmetric AdS$_5$ compactifications.  

\subsection{Ansatz and reduced field equations}
\label{ansatz_section}

Our general Ansatz for the metric and field strength reads as follows 
\bea
ds^2&=& {dr^2 \over L^2-MN} + M dt^2+ 2L dtdx_3  +N dx_3^2+ e^{2V}( dx_1^2 +dx_2^2 ) \\ \no
F&=& B dx_1\wedge dx_2 + E dr\wedge dt + P dx_3\wedge dr
\eea
where $L, M, N, V, E$, and $P$ are functions of $r$ only, and $B$ is a constant as a consequence of
the Bianchi identity.  To identify event horizons, it is useful to write the metric in the form 
\bea\label{alt}
ds^2=  {dr^2 \over L^2-MN}  - {L^2-MN \over N}dt^2
+ N \left(dx_3+ {L\over N} dt \right)^2+ e^{2V}( dx_1^2 +dx_2^2 )
\eea
This Ansatz differs in its choice of radial coordinate from that used in our previous 
papers \cite{D'Hoker:2009mm,D'Hoker:2009bc,D'Hoker:2010rz}.  It was motivated by 
previous work on three dimensional gravity \cite{Clement:1993kc,Banados:2005da}, 
and its advantages will be become clear as we proceed. 

\sm

The metric and field strength are form invariant under the following  coordinate transformations:
\begin{enumerate}
\item $SL(2,R)$: 
\bea
\left(\matrix{t \cr x_3}\right) ~\rightarrow ~\left(\matrix{a & b \cr c& d}\right)
\left (\matrix{t \cr x_3}\right)~,\quad\quad  ad-bc=1
\eea
\item Scale transformations:
\bea
r \rt \lambda^2 r ~,\quad t \rt {1\over \lambda }t~,\quad  x_3 \rt {1\over \lambda }x_3 
\eea
\item $r$ translations:
\bea
r \rt r +  r_0 
\eea
\item Rescaling of $x_{1,2}$:
\bea
x_{1,2} \rt \kappa x_{1,2}
\eea
\end{enumerate}

The reduced field equations are 
\bea\label{baneqs}
M1&&\quad  \left((NE+LP)e^{2V}\right)'+2kBP =0 
\\ \no
M2 && \quad    \left((LE+MP)e^{2V}\right)'-2kBE =0 
\\ \no
E1&& \quad L''+2V'L' +4(V'' +V'^2)L -4PE=0  
\\ \no
E2&& \quad M'' +2V'M'+4(V'' +V'^2)M +4E^2=0 
\\ \no
E3&& \quad N''+2V'N'+4(V'' +V'^2)N+4P^2  =0  
\\ \no
E4 && \quad 6V''f + 8 (V')^2 f + 2 V' f' -g + 4 B^2 e^{-4V}=0
\\ \no
CON && \quad (V')^2 f + V' f' -6 +{g \over 4} + B^2 e^{-4V} 
+ MP^2 + 2 LEP + NE^2 =0 
\eea
where we have defined the following $SL(2,R)$ invariant bilinears
\bea
\label{fg}
f &=& L^2 - MN \\ \no
g&=& (L')^2-M'N'
\eea
There is some redundancy in this set of equations: by taking the derivative of the 
constraint equation
one can show that one linear combination of the equations is satisfied identically.  
Thus, one could omit one of the second order Einstein equations without 
loss of information. 

\subsection{First integrals}

Eliminating $E, P$ and $B$ between the Einstein equations, we find the 
following simple result,
\bea
\label{fV}
\left ( f e^{2V} \right )'' = 24 e^{2V}
\eea
Therefore, we can, in complete generality, solve for $f$ in terms of $V$ as,
\bea\label{fsol}
f(r) = 24e^{-2V(r)} \int_{r_1}^r dr' \int_{r_2}^{r'} dr'' e^{2V(r'')}
\eea
Introducing the potentials $A$ and $C$ respectively for $E=A'$ and $P=C'$,
we may integrate both Maxwell's equations once, 
\bea
\label{maxint}
(NE+LP)e^{2V} + 2 kB C & = & 0
\no \\
(LE+MP)e^{2V} - 2 kB A & = & 0
\eea
The integration constants that arise here have been absorbed into the 
definition of the functions $A$ and $C$. Forming the combinations 
$M \times E1 - L \times E2$, $N \times E2 - M \times E3$, $L \times E3 - N \times E1$,
and using (\ref{maxint}), we find three first integrals of motion, 
\bea
\label{einstint}
\lambda e^{2V} - 4 kB AC = \lambda _0 & \hskip 1in & 2 \lambda = NM' - MN'
\no \\
\mu e^{2V} + 4 kB A^2 = \mu _0 & \hskip 1in & ~ \mu = LM' - ML'
\no \\
\nu e^{2V} + 4 kB C^2 = \nu _0 & \hskip 1in & ~ \nu = NL' - LN'
\eea
Here, the combinations $\lambda, \mu, \nu$ are the analogues of angular momentum
components for SL(2,R) and, just as $L,M,N$ do, they transform under the vector 
representation of SL(2,R). They satisfy the following purely kinematic relation, 
\bea
\label{radial}
(f')^2 = 4 fg + 4 (\lambda ^2 - \mu \nu)
\eea
Finally, $\lambda _0, \mu _0$, and $\nu_0$ are the constant values of the 
first integrals.

\subsection{Asymptotically AdS$_5$ solutions}

In these coordinates, the pure AdS$_5$ solution takes the form
\bea
ds^2 &=& {dr^2 \over 4 r^2} +2r (-  dt^2 +  dx_3^2 + dx_1^2 +dx_2^2)  \\ \no
F&=& 0
\eea
More generally, we will be looking for asymptotically AdS$_5$ solutions, for which the leading
behavior at large $r$ is given by, 
\bea
\label{asmet}
ds^2&\sim & {dr^2 \over 4r^2} +c_M  rdt^2  +c_Nr dx_3^2+  2c_L r dx_3 dt+  c_V r (dx_1^2 + dx_2^2) \\ \no
F&\sim& b dx_1 \wedge dx_2 +  {c_E\over r^2} dr \wedge dt +   {c_P\over r^2} dx_3 \wedge dr
\eea
with $c_L^2 - c_M c_N=4$, as implied by the field equations. Here, we denote the magnetic 
field $b$, reserving use of $B$ for the case where $x_{1,2}$ are scaled to put the metric
in canonical form. 

\sm

By using the coordinate transformations discussed above, we have some freedom to  change  
the asymptotic parameters.  For instance, we can use the $SL(2,R)$ and scale transformations
to set $c_{L,M,N}$ to any desired values
obeying $c_L^2 - c_M c_N=4$, and also choose  an arbitrary value for $c_E$, assuming that it is nonzero
to begin with.    The freedom to translate $r$ can then be used to set the location
of the event horizon, if present, at $r=0$. 

\sm

With $b=0$, a  familiar asymptotically AdS$_5$ solution is the electrically charged 
Reissner-Nordstrom solution, with a near horizon AdS$_2 \times R^3$ geometry.    
In this work we are interested in
solutions with $b\neq 0$, corresponding to a gauge theory with a nonzero background magnetic field.

\subsection{Near horizon solutions}
\label{nhsolutions1}

With $b\neq 0$, it is possible to find exact solutions where $V$ is constant, so that the metric
takes the form $M_3 \times R^2$.   Such solutions can, and will,  serve as near horizon geometries 
for full asymptotically AdS$_5$ solutions.   In fact, it is not hard to find the {\it general} solution
with constant $V$.   As shown in Appendix \ref{nhsolutions}, for $k\neq 1$ (recall that we have 
restricted to $k \geq 0$), the general solution with constant $V$ is given by, up to coordinate transformations,\footnote{The $\eta =1$ solution obviously breaks down for 
$k=1 /2$; this value of $k$ is one, among several, ``special" values of the Chern-Simons coupling $k$.}
\bea\label{nhsol}
F&=& b dx_1 \wedge dx_2 +q r^{\eta k-1} dr \wedge dt  \\ \no
ds^2 & = & {dr^2 \over 4 b^2 r^2 }- \left(\beta+\alpha r  
+{ q^2 \over k (k-{1 \over 2}\eta )}r^{2\eta k}\right) dt^2 +4\eta br dt dx_3 
+ {b\over \sqrt{3}}(dx_1^2 +dx_2^2) 
\eea
with $\eta = \pm 1$. 
This solution appears to be new, although some aspects are familiar.   If we set $q=0$
the solution is identified as AdS$_3 \times R^2$ which is the near horizon part of the magnetic
brane solutions found in \cite{D'Hoker:2009mm}.    The AdS$_3$ part can equally well be taken
to be a nonextremal, rotating BTZ solution, since these can be obtained from AdS$_3$ by applying
the coordinate transformations discussed in section \ref{ansatz_section}.  Note also that we
can use the residual coordinate freedom to set $b, q$ and $\alpha$  to convenient values, 
such as $b=\sqrt{3}$, $q=1$ (assuming nonzero $q$), and $\alpha=0$.  

\sm

A simple  but important  fact is that the $\alpha=\beta=q=0$ solution is invariant under the following coordinate
transformations, which are a combination of scale and $SL(2,R)$ transformations, 
\bea\label{scale}
r \rt \lambda r~,\quad t \rt  {1\over \lambda^w}t~,\quad x_3 \rt  {1\over \lambda^{1-w}}x_3
\eea
for {\it any}  parameter $w$. If we write $t \sim  (x_3)^z$ we have
\bea\label{zw}
z = {w \over 1-w} 
\eea
Eventually, $z$ will  be identified with a dynamical critical exponent.  Of course, nothing we have
said so far determines any particular value of $w$ and hence of $z$.  Preferred values emerge once 
we match the near horizon geometry  on to a  low temperature asymptotically AdS$_5$ solution. 
Doing so, we will find that $w=1/2$ and $w=3/4$ are realized, leading to $z=1$ and $z=3$ for
the values of the dynamical critical exponent in the dual field theory. 

\sm 

For nonzero $q$, the $(r,t,x_3)$ part of the metric coincides with the corresponding three
dimensional part of the ``Schrodinger" spacetimes, proposed in 
\cite{Son:2008ye,Balasubramanian:2008dm} in connection with non-relativistic 
AdS/CFT.\footnote{In those solutions there are additional spatial directions appearing 
in the metric as $r dy_i dy_i$, and associated
Galilean boost symmetries.}  The nonzero $q$ solutions, but still with $\alpha=\beta=0$, are invariant under 
\bea\label{scale2}
 r \rt \lambda r~,\quad t \rt {1\over \lambda^{\eta k}}t~, \quad x_3 \rt {1\over \lambda^{1-\eta k} }x_3
\eea
Perhaps surprisingly, this scale transformation will end up playing  no role in our
discussion of scaling near quantum critical points; rather, as noted above, it is the scale transformations 
(\ref{scale}) with $w=1/2$ and  $3/4$ that turn out to be important.\footnote{In fact, for $k\leq 3/4$ the scale 
transformation (\ref{scale2}) {\it does} determine the dynamical critical exponent \cite{inprog}.  The importance of $k=3/4$ can be seen from the discussion in section 4.3.6.}  Indeed, note that since the
supersymmetric value of $k$ is greater than $1$, the dynamical critical exponent corresponding to 
(\ref{scale2}) would be {\it negative} for the supersymmetric case.   

\sm

Solutions with the same three dimensional metric,  but with the time and space coordinates swapped, have been studied recently as solutions of topologically massive gravity \cite{Anninos:2010pm}.   By virtue of their asymptotic symmetry algebra, they were found to possess a chiral Virasoro algebra.  

\sm

Precisely at $k= 1$ a new family of solutions emerges.  These are the warped AdS$_3 \times R^2$
solutions found in \cite{D'Hoker:2009bc}, whose three dimensional part was studied previously
in the context of topologically massive gravity \cite{Anninos:2008fx}.   These solutions will not play any
role in what follows, but we present them here for completeness.   The solutions are
\bea\label{kone}
ds^2 & = & {dr^2 \over 12 r (r+r_0)} -2 r(q^2 r+ {6 r_0})dt^2 + 4br dt dx_3 +dx_3^2 + dx_1^2 + dx_2^2  \\ \no
F&=& b dx_1 \wedge dx_2 + q dr \wedge dt  
\eea
with $q^2 +2b^2 =6$.

\subsection{Physical parameters}

Given an asymptotically AdS$_5$ black hole solution we will be interested in computing the magnetic
field,  charge density,  temperature and entropy density.    These quantities can be read off from the asymptotic
data near the horizon and at infinity. 

\sm

  For the near horizon behavior it
is clearest to work with the metric in the form (\ref{alt}).  The horizon occurs when $L^2-MN=0$, which
we assume is satisfied at $r=0$.   Finite entropy density solutions will have $N(0)\neq 0$.   The value of $L$ at 
the horizon is interpreted as a chemical potential conjugate to $P_3$, the momentum along $x_3$. 
This follows from the fact that the existence of a smooth Euclidean section requires that periodic
shifts of imaginary time be accompanied by a shift of $x_3$ proportional to $L(0)$.  We will
restrict attention to solutions in which the only thermodynamical potentials are temperature and 
a chemical potential for the charge density, and hence we demand $L(0)=0$.  For the solutions
we find, $L$ will vanish linearly in $r$.    It then follows that at a finite entropy horizon we require
$M$ to vanish; $M'(0)$ will be proportional to the Hawking temperature.  

\sm

We assume a metric with asymptotic behavior (\ref{asmet}).  To read off physical quantities it
is convenient to perform coordinate transformations to put the asymptotic metric in canonical form.
The  $SL(2,R)$ transformation 
\bea
t \rt t - {c_L \over c_M}x_3  
\eea
removes the $dx_3 dt$ cross term from the metric, i.e. sets $c_L=0$, while preserving 
the condition $L(0)=0$.     Next, we can perform
scale and $SL(2,R)$  transformations to set $-c_{M}=c_{N}=c_{V}=2$.   

\sm

After applying these transformations, which of course also act on the near horizon part of the metric,
it is straightforward to work out expressions for the physical parameters.  We then express the results in
terms of the asymptotic coefficients appearing in the general solution, with arbitrary $c_{M}, c_{N}, \ldots$.   
The relevant formulas are
\bea\label{physpar} 
 B&=& {2b\over c_V} \\ \no
 \rho & =& 4   \sqrt{2 \over -c_M} c_E\\ \no
 s &=&{1\over 4}e^{2V(0)} \sqrt{N(0)}\sqrt{-c_M \over 2} \left({2 \over c_V}\right)   \\ \no 
T &=&{1 \over 4\pi}  \sqrt{2 \over -c_M}\left( -M'(0) \sqrt{N(0)}\right) 
\eea
Here $s$ is the entropy density with Newton's constant scaled out: $s = G_5 S/{\rm Vol}$.  
The charge density $\rho$ is determined from the boundary current according to the standard
AdS/CFT dictionary; see \cite{D'Hoker:2009bc} for details. 

\sm

These physical parameters are dimensionful, and can be rescaled by a coordinate transformation
that preserves the asymptotic AdS$_5$ metric.  It is thus only dimensionless quantities that are
meaningful, and these are defined as
\bea\label{dimless}
\Bh &\equiv & {B \over \rho^{2/3}}   \\ \no
\sh &\equiv & {s \over B^{3/2}}   \\ \no
\Th &\equiv & {T\over B^{1/2} }   \\ \no
\eea
Here we are defining $\sh$ and $\Th$ in a slightly more convenient manner than in \cite{D'Hoker:2010rz},
where in the denominator we used the combination $B^3 +\rho^2$; this will clearly have no effect
on the scaling properties since we'll always have nonzero $B$.

\section{Zero temperature asymptotically AdS$_5$ solutions}
\setcounter{equation}{0}
\label{ZeroT}

We now turn to the construction of asymptotically AdS$_5$ solutions representing 
zero temperature, finite density, matter in a magnetic field.   As we will see, such solutions can be
found by quadrature in terms of a single universal $V$ function; $V$ is ``universal" in the sense
that it has no dependence on the magnetic field, charge density, or Chern-Simons coupling $k$, and hence can 
be determined (numerically) once and for all.  

\subsection{Pure magnetic solutions}

We begin by recapitulating the solutions obtained in \cite{D'Hoker:2009mm}, with a magnetic
field but with vanishing electric charge.    We can choose coordinates such that these solutions obey
\bea
M = N  =E=P=0
\eea
Furthermore, by scaling $x_{1,2}$ we can set 
\bea
b= \sqrt{3} 
\eea
The Ansatz is then
\bea
\label{AdS3}
ds^2 & = & {dr^2 \over L^2} +2 L dt dx_3 + e^{2V}( dx_1^2 + dx_2^2 )
\\ \no
F&=& \sqrt{3} dx_1 \wedge dx_2 
\eea
so that $t$ and $x_3$ play the role of lightcone coordinates.  

\sm

The problem is reduced  to solving for $L$ and $V$, which obey
\bea\label{LVeqtns}
e1&& \quad L''+2V'L' +4(V''+V'^2)L  =0 \\ \no
e4 && \quad {3\over 2}L^2V''+{1\over 2}e^{-4V}  (L^2 e^{4V})'V' -{1\over 4}L'^2+3e^{-4V}=0 \\ \no
con && \quad e^{-V}  (L^2 e^V) 'V'+{1\over 4}L'^2 +3e^{-4V}-6=0 
\eea
From (\ref{fsol}) we  obtain $L$ in terms of $V$.  We choose the boundary conditions 
$L(0)=L'(0)=0$, corresponding to a zero temperature horizon at $r=0$.  Then,
\bea\label{fsol1}
L(r)^2 = 24e^{-2V(r)} \int_{0}^r dr' \int_{0}^{r'} dr'' e^{2V(r'')}
\eea
and the problem is reduced to finding $V$.   

\sm

\begin{figure}[h]
\begin{centering}
\includegraphics[scale=0.6]{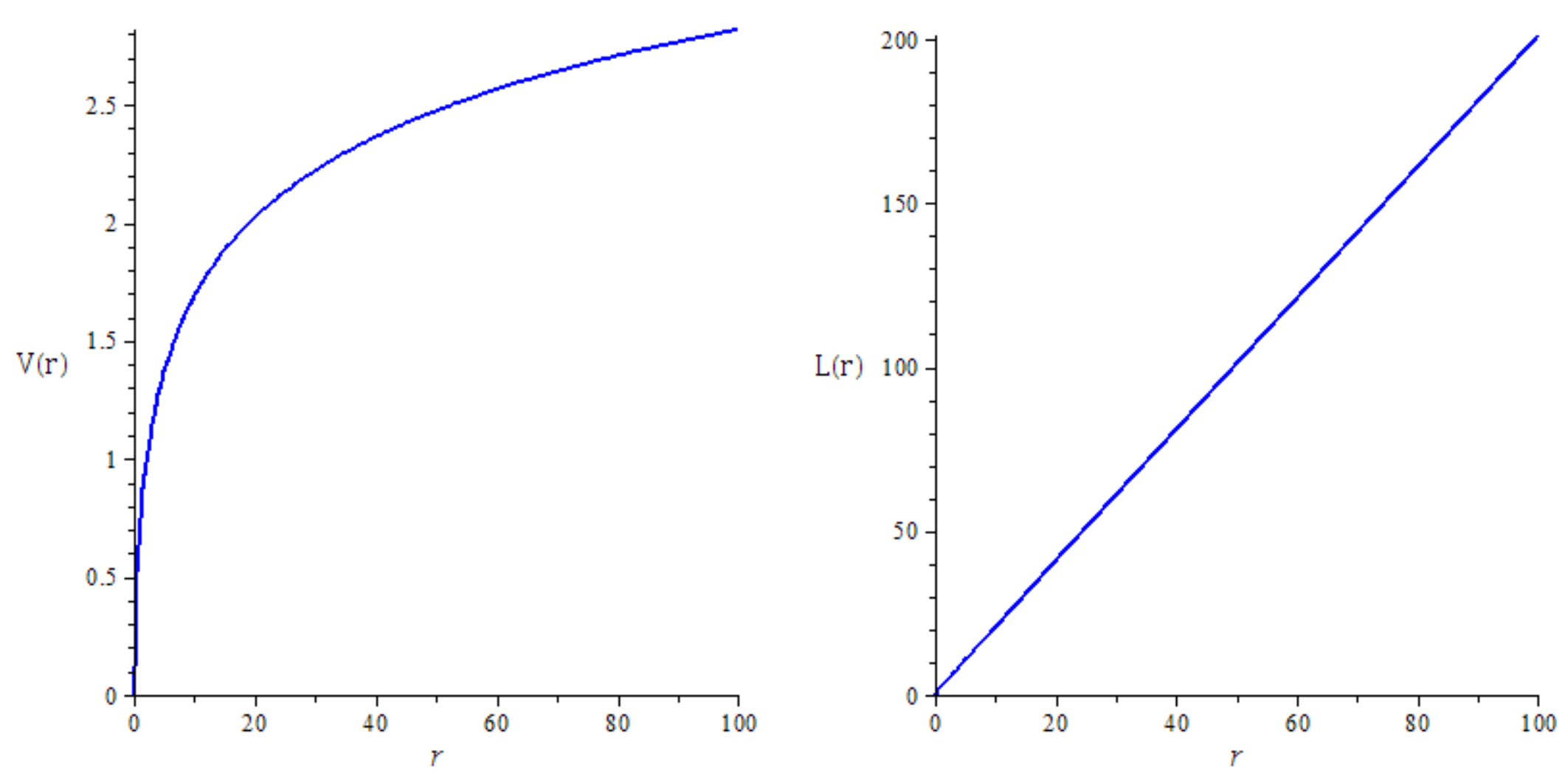}
\caption{Numerical solution of (\ref{LVeqtns})  for $V(r)$ and $L(r)$ subject to boundary conditions (\ref{VLas}) }
\label{LVfig}
\end{centering}
\end{figure}

As in  \cite{D'Hoker:2009mm} we are interested in solutions that interpolate between 
AdS$_3 \times R^2$ at small $r$ and AdS$_5$ at large $r$.  
At small $r$ the equations admit a solution of the form
\bea 
V(r) = v_1 r^\sigma + v_2 r^{2\sigma} + v_3 r^{3\sigma} + \cdots
\eea
with $\sigma$ obeying the quadratic equation 
\bea
3\sigma^2 +3\sigma -4=0
\eea
We  choose the root such that $V$ is finite at $r=0$, and so take
\bea
\label{sigma}
\sigma = {\sqrt{57} \over 6}-{1\over 2} \approx  .758
\eea
 Using our freedom to perform a scale transformation on $(r,t,x_3)$ we set $v_1=1$, and then
find for the first few terms at small $r$:
\bea\label{VLas}
V(r)  &=& r^\sigma - 3 {(2\sigma+1)(3\sigma-2) \over 3\sigma+1} r^{2\sigma} + \cdots \\ \no
L(r) &=& \sqrt{12} r \left(1 - 2 {2\sigma -1 \over \sigma+1} r^\sigma + \cdots \right)
\eea
Followed out to large $r$ these initial data match on to an asymptotically AdS$_5$ solution.  We have
not succeeded in solving for $V$ analytically, but numerical integration is straightforward.  The
resulting functions are shown in Fig. \ref{LVfig}. 
The large $r$ asymptotics are
\bea\label{magas}
e^{2V(r)}  &\sim &  c_V r~,\quad\quad c_V \approx 2.797 \\ \no
L(r) & \sim & 2r 
\eea 

The physical interpretation  of this solution was explored in  \cite{D'Hoker:2009mm}.  
In the dual field theory it corresponds to massless fermions in a magnetic field.  The modes
occupying the lowest Landau level give rise at low energies to a $D=1+1$ dimensional CFT,
which accounts for the existence of the near horizon AdS$_3$ factor.  In the case of
${\cal N}=4$ SYM, the free field central charge was compared with the Brown-Henneaux
central charge, and found to differ by a factor of $\sqrt{3/4}$.  Exact agreement is not expected
since these solutions are non-supersymmetric.  

\subsection{Charged solutions}

We now generalize to include a nonzero electric charge density.  Rather remarkably, we find
that the full solution can be solved by quadrature in terms of the same $V$ function as appeared
in the pure magnetic solution above.   This fact allows us to  deduce analytically  all the most interesting
physical properties of these solutions.  These properties include the existence of a critical
electric charge density at which the system undergoes a quantum phase transition, and expressions
for the low temperature thermodynamics at and  away from the critical point. 

We describe these solutions in coordinates such that $N=P=0$.  Examining the field equations, 
we see that   $L$ and $V$ obey precisely the same equations as in the pure magnetic case, and
therefore we can simply carry over the results from that solution.   To complete the solution we need
to solve for $E$ and $M$.

\subsubsection{Solving for $E$} 

Recalling that we are scaling $x_{1,2}$ to set $b=\sqrt{3}$,   equation M2 now reads
\bea
 (L e^{2V}E)'=\sqrt{12}kE
\eea
which we integrate as
\bea
E(r) =  {2  c_{V} c_{E} \over L(r) e^{2V(r)} } \exp \left \{ \sqrt{12}k \int ^r _\infty { dr'
\over L(r') e^{2V(r')}} \right \}
\eea
$c_E$ is an integration constant, while $c_V$ is the same number that appeared in (\ref{magas}). 
It will be convenient to define the following function,
\bea
\psi (r) \equiv \int ^r _\infty { dr' \over L(r') e^{2V(r')}}
\eea
in terms of which the  gauge potential, defined as $E(r)=A'(r)$
with $A(0)=0$, is given by, 
\bea
\label{Azero}
A(r) = {c_{V} c_{E} \over \sqrt{3}k} e^{\sqrt{12}k \psi(r)}
\eea

\sm

The asymptotics of $\psi$ are found to be, 
\bea
r \to 0 & \hskip 0.3in & \psi (r) \sim {\ln r \over \sqrt{12} } 
+ {1 \over \sqrt{12} k} \ln \left ( { \sqrt{3} e_0 \over   c_{V}c_{E}} \right )
\no \\
r \to \infty && \psi (r) \sim  - {1 \over 2  c_{V} r} 
\eea
where $e_0$ in the last term of the first line arises from the regularized integral,
\bea
\label{ezero}
e_0 = { c_{V} c_{E} \over \sqrt{3}} \exp  \left \{ \sqrt{12}k \int ^0 _\infty dr' \left [
{1 \over L(r') e^{2V(r')}} - { 1 \over \sqrt{12} r'(r'+1)} \right ] \right \}
\eea
This  integral is convergent and produces
the following asymptotics of $E(r)$,
\bea
\label{Eas}
r \to 0 & & E(r) \sim e_0 \, r^{k-1}
\no \\
r \to \infty && E(r) \sim {c_{E} \over r^2} 
\eea
The asymptotics of $\psi$  guarantee that $A$, as defined in (\ref{Azero}), automatically 
vanishes at the horizon, as long as $k>0$. Note that the ratio $e_0/c_{E}$ depends only
on the properties of the purely magnetic $T=0$ solution.

\subsubsection{Solving for $M$}

The remaining equation E2 for $M$ may be integrated by noticing that it
is a linear equation in $M$, whose homogeneous part coincides with 
the equation for $L$. Thus, we know that we can integrate by quadrature.
We set
\bea
M(r) = L(r) \f (r)
\eea
and obtain the following equation for $\f$,
\bea
\left ( L^2 e^{2 V} \f' \right )' = - 4 E^2 L e^{2V}
\eea
In terms of $\psi$, we have 
\bea
\left ( L^2 e^{2 V} \f' \right )' = - 16 c_{V}^2 c_{E}^2 \psi ' e^{4\sqrt{3} k \psi }
\eea
It is straightforward to integrate this equation, and we find,
\bea 
L^2 e^{2 V} \f' =  \sqrt{12} \beta -{ 4 c_{V}^2 c_{E}^2 \over \sqrt{3}k }   e^{4\sqrt{3} k \psi (r) } 
\eea
where $\beta$ is an integration constant. Integrating once more, we find, 
\bea\label{phis}
\f (r) = - {\a \over \sqrt{12}} + \sqrt{12} \beta \int _\infty ^r { dr' \over L(r')^2 e^{2V(r')}}
- { 4  c_{V}^2 c_{E}^2 \over \sqrt{3}k} \int _\infty ^r  dr'  { e^{4\sqrt{3}k \psi (r')} \over L(r')^2 e^{2V(r')}}
\eea
with $\a$ another integration constant. 

\sm

It is now straightforward to evaluate the asymptotics of $M$, and we find, 
\bea
\label{Mas}
r \to 0 & \hskip 0.3in & M (r) \sim   -\beta  - \tilde \a r 
- {  e_0^2 r^{2k} \over k(k-{1\over 2})} 
\no \\
r \to \infty && M (r) \sim  - { \a \over \sqrt{3} }  r
\eea
where 
\bea
\label{alf}
\a - \tilde \a = 16 c_V^2 c_E^2 J(k) \hskip 1in 
J(k) = {1 \over 2k} \int _0 ^ \infty dr' { e^{4\sqrt{3} k \psi (r')} \over L(r')^2 e^{2V(r')}}
\eea
The integral $J(k)$ is convergent for all $k >1/2$ and positive. 

\subsection{Interpretation and emergence of the critical magnetic field}
\label{critmag}

We are interested in  solutions that represent zero temperature limits of black hole solutions,
with a horizon at $r=0$.   We therefore demand that $M$ vanish at $r=0$, which requires that we 
set $\beta=0$.\footnote{In fact, the $\beta \neq 0$ solutions also have an interpretation as
the zero temperature limit   of  black hole solutions, but they necessarily involve a non-zero chemical
potential conjugate to momentum along $x_3$.  This follows since for nonzero $\beta$ we can
always shift $r$ such that $M(0)=0$, but then $L(0)\neq 0$, which is equivalent to having such
a chemical potential.   It would be interesting to explore these solutions further (the neutral version is
studied in Appendix D of  \cite{D'Hoker:2009bc}), but here we exclude
them since we are working at vanishing $P_3$ chemical potential.  }   From the expression in 
(\ref{phis}) it is then apparent that 
\bea
-{ \a \over \sqrt{12} }  \leq { M (r) \over L(r) } \leq - { \tilde \a \over \sqrt{12}}
\eea
uniformly throughout $0\leq r < \infty$. Since $L(r)$ is positive, 
$M(r)$ will be negative definite as long as $\tilde \alpha$ is non-negative. 
Finite temperature black hole solutions will have $M(0)=0$ with $M'(0) <0$, the Hawking 
temperature being proportional to $-M'(0)$.  
Thus, we need $\tilde \alpha \geq 0$  in order for our solution to represent 
the zero temperature limit of such black holes.  In view of (\ref{alf}),  we thus require
\bea
\label{bound}
\alpha \geq 16 c_V^2 c_E^2 J(k) 
\eea
This translates into a bound on the dimensionless magnetic field $\Bh$.    
From (\ref{physpar}), (\ref{dimless}),  and (\ref{Mas}), we have 
\bea\label{Bhat}
\Bh = \left(3\over 4\right)^{1/3} { \alpha^{1/3} \over c_V c_{E}^{2/3}}
\eea
and so that the bound (\ref{bound}) translates to $\Bh \geq \Bh_c$ with the critical field 
given as a function of the universal number $c_V$, and the Chern-Simons coupling $k$ only,   
\bea\label{Bhatc}
 \Bh_c = \left( 12 J(k) \over c_{V} \right)^{1/3} 
\eea
Numerical integration yields the  values shown in Table 1 for $\Bh_c$ for various values of $k$.
\begin{table}[htdp]
\begin{center}
\begin{tabular}{|c|c|} \hline
$k$ & $\Bh_c $ \\ \hline \hline
 .50001&  35.4050648722\\ \hline 
.6 & 1.44700934549  \\ \hline
.75 &.916107730288  \\ \hline 
1 &.600520361557  \\ \hline 
${2/ \sqrt{3}}$ &.499424265324  \\ \hline 
2 &.264993652464  \\ \hline 
5 &.101365592402  \\ \hline 
10 & .050219317885  \\ \hline 
\end{tabular}
\end{center}
\caption{ The critical value  of the dimensionless magnetic field to charge density ratio 
for selected values of $k$. }
\end{table}
As can be seen from the integral representation, $\Bh_c$ diverges at $k=1/2$ and goes
to zero at large $k$.   For the supersymmetric value, $k=2/\sqrt{3}$, we recover the critical
value that arose in the numerical studies conducted in \cite{D'Hoker:2010rz}.   The interpretation
of this critical field in terms of the dual gauge theory is of course an interesting question, on which 
we will comment  in the discussion section.

\sm

Before turning to finite temperature, we pause to note a potentially confusing point regarding
these zero temperature solutions.  After setting $b=\sqrt{3}$ we presented the solution
in terms of two integration constants, $c_E$ and $\alpha$, and showed how they determined the
value of $\Bh_c$.   It is easy to see however that, up to coordinate transformations,  solutions with
different values of $(c_E,\alpha)$ are in fact equivalent.  In particular, the $SL(2,R)$
transformation $x_3 \rt x_3 + ct$ can be used to shift $\alpha$, and  a rescaling of $t$ and
$x_3$ will rescale $c_E$.    Even though these solutions are coordinate equivalent, it is appropriate
to treat them as physically distinct when one regards them as zero temperature limits of finite
temperature solutions.  At finite temperature, and in particular when $N(0)\neq 0$, solutions with different $\Bh_c$ are {\it not} related
by an allowed coordinate transformation: we are not allowed to perform the  $SL(2,R)$
transformation $x_3 \rt x_3 + ct$ due to the condition that $L$ vanish at the horizon, which in 
turn is mandated by the absence of a chemical potential for $P_3$.   In order to have a continuous
zero temperature limit, we need to keep the full family of zero temperature solutions. 
We will return to this topic in section \ref{bqcrit} where we discuss the approach to criticality in
the parameters $(b,q)$.

\section{Low temperature thermodynamics}
\setcounter{equation}{0}
\label{LowT}

The solutions constructed in the last section carry nonzero charge density and magnetic field, 
but have vanishing temperature and entropy density.  We now want to heat them up and study the
low temperature behavior of the entropy density.    

\subsection{Matched asymptotic expansions} 

Since analytic solutions at arbitrary $T$ are not available, we need to proceed perturbatively in small $T$.
However, straightforward perturbation theory around the zero temperature solutions will not work,
as the perturbations diverge at the horizon; at the horizon, the change from  zero to finite temperature
is not a small perturbation.  Instead, we need to employ a matched asymptotic expansion.  The
basic idea is well illustrated by a simple example.

\sm

Below is the metric for finite temperature D3-branes in asymptotically flat space:
\bea\label{D3}
ds^2&=& H^{-{1\over 2}}(-fdt^2+ dx^i dx^i) + H^{{1\over 2} }f^{-1} dr^2 + H^{1/2}r^2 d\Omega_5^2 
\\ \no
H&=& 1+ {L^4\over r^4}~,\quad  f = 1- {r_+^4 \over r^4} 
\eea
To make the analogy with our problem, suppose we only knew the solution (\ref{D3}) in two
limiting cases: 1)~  $r_+=0$, which is the zero temperature solution;  ~2)~ $r\ll L$, corresponding
to omitting the $1$ in $H$, which yields the AdS$_5$ Schwarzschild solution times the  5-sphere. 
Using perturbation theory, how could we construct the full solution in the regime $r_+ \ll L$ 
(the low temperature regime) and extract
the low temperature thermodynamics?   Note that  from the knowledge of the AdS Schwarzschild
solution one can of course work out its entropy density and temperature, but to import this result to the
asymptotically flat solution one needs additional information, since one does not know {\it a priori}
how the AdS Schwarzschild time coordinate matches onto the time coordinate in the asymptotically
flat region.  They could differ by a scale transformation, which would rescale the temperature; this is a
key point for understanding the low temperature behavior of the solutions studied in this 
paper.\footnote{For the D3-brane example, due to the boost invariance of the zero temperature 
solution there is actually no ambiguity in determining $s(T)$ in this manner, but since boost 
invariance is absent for
our solutions, let us ignore this fact to maintain a faithful analogy.} 

\sm

 In the matched asymptotic expansion approach we consider two different perturbation problems. 
In the first, we perturb around the AdS Schwarzschild solution, requiring smoothness at the horizon,
which leads to perturbations that grow with $r$.  In the second, we perturb around the zero
temperature solution, maintaining asymptotic flatness, which gives perturbations that grow at small $r$.
If $r_+ \ll L$, there is a parametrically large overlap region, $r_+ \ll r \ll L$ where both expansions
are valid.  In this region we can match up  the free parameters appearing in the perturbations, and
so connect near horizon data with that at infinity.  For the D3-brane example, the perturbed solution
in the overlap region is given by 
\bea 
 ds^2 = ds_{{\rm AdS}_5 \times S^5}^2  -{r^6\over 2L^6}(-dt^2 + dx_i^2)+{r_+^4\over r^2}dt^2 
 +{r^2\over 2L^2} dr^2  +{r_+^4\over r^6}dr^2
\eea
illustrating the presence of perturbations that grow at small and large $r$.  

\sm

Our problem is similar, with a few extra complications.   First, we have two independent perturbations,
corresponding to changing $\Th$ and $\Bh$.   Second, our zero temperature solutions are not
given by a closed form expression.   But it is still possible to use this method to determine
the low temperature thermodynamics.

\subsection{Summary of matching computations}

We defer a detailed solution of the perturbations problem to the next section.  Here we just
summarize the basic results that are needed to determine the thermodynamics.  
The first part of the problem consists of expanding around a BTZ solution, and the second
in expanding around the zero temperature asymptotically AdS$_5$ solutions constructed above.
These expansions are matched in an intermediate region where both solutions approach
AdS$_3 \times R^2$ plus small perturbations.  This AdS$_3 \times R^2$ region is the
analog of AdS$_5 \times S^5$ in the D3-brane example.   Linearizing around AdS$_3 \times R^2$
yields the following result
\bea
\label{overlapb}
E(r) & = & e_0 r^{k-1}
\no \\
P(r) & = & p_0 r^{-k-1}
\no \\
L(r) & = & l_0 + 2 b r  -4\sqrt{3}\left(2\sigma -1 \over \sigma+1\right) v_+ r^{\sigma+1}
 -4\sqrt{3}\left(2\sigma +3 \over \sigma\right) v_- r^{-\sigma}
\no \\
M(r) & = & m_0 + m_1 r
\no \\
N(r) & = & n_0 + n_1 r
\no \\
V (r) & = & v_+ r^\sigma + v_- r^{-\sigma -1}
\eea 
where all free integration constants are indicated and $b=\sqrt{3}$.  The problem consists of relating these integration
constants to the parameters that appear in the BTZ and asymptotically AdS$_5$ perturbation problems. 

\subsubsection{Perturbations around BTZ}

An exact solution to the field equations is given by BTZ$\times R^2$ 
with magnetic flux, which we write as
\bea
\label{BTZ1}
ds^2&=& {dr^2 \over 12r^2 + mn r} -m r dt^2 
+4\sqrt{3}rdt dx_3 +  n dx_3^2 +dx_1^2 +dx_2^2 \\ \no
F&=& \sqrt{3} dx_1 \wedge dx_2
\eea
{\it i.e. } $b=\sqrt{3}$ and, 
\bea
\label{BTZ2}
 L =2 \sqrt{3}r~,\qquad M=-mr~,\qquad N=n~,\qquad V= E=P=0
\eea
In order for this solution to match onto a charged asymptotically AdS$_5$ solution,
we need to add a two parameter family of perturbations.  One parameter corresponds
to adding electric charge, and the second to inducing a flow towards AdS$_5$.    The perturbation parameters
are $q$ and $v_0$, given by the values of $E$ and $V$ at the horizon,\footnote{Note that
the normalization for the electric charge at the horizon $q$ used in \cite{D'Hoker:2010rz} was
chosen with $n=1$, and differs by a factor of $\sqrt{n}$ from the normalization used here.}
\bea
E(0)=q \hskip 1in  V(0)=v_0
\eea
A smooth perturbation  with these boundary conditions can be found analytically in terms of
hypergeometric functions.  For present purposes we just need results for $E$ and $V$.  At large
$r$, their asymptotics match onto (\ref{overlapb}) with parameters
\bea
e_0 =  {E_k q \over (mn)^{k-1}}~,\quad\quad  v_+ =  {V_\sigma  \over (mn)^\sigma} (v_0  + nq^2 A_q)
\eea
and where
\bea
E_k=   {12^{k-1} \Gamma(2k) \over k \Gamma(k)^2}~,\quad\quad 
V_\sigma =   { 12^\sigma \Gamma(1+2\sigma) \over \Gamma(1+\sigma)^2  }
\eea
Here, the coefficient $A_q$ depends only on the $k$, but not on $q,m$ or $n$.
Its construction will be given in  (\ref{AV1}) of section \ref{match1}.

We now match this result to the small $r$ asymptotics of the asymptotically AdS$_5$ solution. Here
we note that $e_0$ and $v_+$ are nonzero already in the unperturbed solution, and so to leading
order we need not concern ourselves with the corrections.   For $E$ we already defined $e_0$
to yield the small $r$ asymptotics (see  (\ref{Eas})) , and for $V$ we chose $v_+=1$ (see (\ref{VLas})).  Hence the matching 
yields
\bea
\label{match7}
 {E_k q \over (mn)^{k-1}} = e_0~,\quad\quad
   {V_\sigma  \over (mn)^\sigma} (v_0 + nq^2 A_q)=1 
\eea

\subsubsection{Perturbations around asymptotically AdS$_5$ solution}

Now consider perturbations around the zero temperature, asymptotically AdS$_5$ solution. 
We demand that these perturbations preserve the conformal boundary metric, meaning that they
leave the asymptotic metric  constants $c_{L,M,N,V}$ unchanged.    With these parameters
fixed, and with the magnetic field fixed at $b=\sqrt{3}$, it is apparent from (\ref{physpar})
that any change in  $\Bh$ occurs via a change in $c_E$.  

\sm

Up to coordinate transformations, there is a two parameter family of such perturbations, 
corresponding to changing $\Th$ and $\Bh$; we denote the corresponding  perturbation 
parameters as $\eps_T$ and $\eps_B$.  After turning on these parameters, we extract
the small $r$ asymptotics and read off the parameters in (\ref{overlapb}).  Here
we only need the results for $M$ and $N$, which are 
\bea
m_0= 0~,\quad\quad m_1 = - \left ( \tilde \a + \eps_T \right )~,\quad\quad n_0 =  C_T \eps_T + \eps_B ~,\quad\quad n_1=0
\eea
We also need to know the change in the large $r$ asymptotics of the electric field, to keep
track of how the charge density changes:
\bea
r \rt\infty \hskip 0.7in  E \sim {c_E  \over r^2}~,\quad\quad  c_E =  c_{E0}+ +C_B \eps_B
\eea
where we now denote the electric field coefficient  appearing in the unperturbed solution
by  $c_{E0}$. The coefficients  $C_T$ and $C_B$ depend on the Chern-Simons coupling 
$k$, and their values may be extracted from the explicit form of the perturbations
around the asymptotically $AdS_5$ solution. They will be computed analytically in terms 
of these data in section 5. 

\sm

Since $M$ and $N$ are already nonzero in the unperturbed BTZ solution  we can match
to their values, neglecting the higher order corrections.  This yields 
\bea\label{asmatch}
  \tilde \a   +\eps_T = m~,\quad\quad     C_T \eps_T + \eps_B  = n
\eea
where we recall that $\a - \tilde \a =  16  c_{V}^2 c_{E0}^2  J(k)$.

\subsection{Thermodynamics and scaling}

We now have all the information we need to read off the low temperature thermodynamics.
The results are different depending on whether we are at or above the critical magnetic field,
and so we treat these cases separately.

\subsubsection{Low temperature thermodynamics for $\Bh>\Bh_c$}

We first consider sitting at fixed $\Bh > \Bh_c$ and lowering the temperature to zero. 
To stay at fixed $\Bh$ we set $\eps_B=0$, and then (\ref{asmatch}) yields
\bea\label{epsn}
C_T \eps _T = n
\eea
To go to zero temperature we need to take $\eps_T \rt 0$.  The condition $\Bh >\Bh_c$
implies $ \tilde \a   >0$, so then (\ref{asmatch}) gives that $m$ should asymptote  to a finite value 
\bea
m =  \tilde  \a  = \alpha - 16 c_V^2 c_{E0}^2 J(k)
\eea
From (\ref{Bhat}) and (\ref{Bhatc}) this yields
\bea\label{alpham}
{\alpha \over m} = {\Bh^3 \over \Bh^3 -\Bh_c^3}
\eea
Now, using $c_M = -\alpha/\sqrt{3}$ (which follows from (\ref{Mas})), $b=\sqrt{3}$, and the
formulas in (\ref{physpar}), (\ref{dimless}), it is straightforward to evaluate the following
results as $\eps_T \rt 0$:
\bea\label{stforms} 
\sh = {1\over 24} \sqrt{c_V n \alpha}~,\quad\quad  \Th = {1\over 4\pi}{m\sqrt{c_V n }\over \sqrt{\alpha}} 
\eea
Eliminating $m$ and $\a$ in favor of $\hat T$ and $\hat s$ automatically cancels 
the dependence on $n$, and we find the following expression for the leading low 
temperature behavior of the entropy,
\bea
\sh = {\pi \over 6}\left( {\Bh^3 \over \Bh^3 -\Bh_c^3}\right) \Th 
\eea
as announced in (\ref{entropy1}) of the introduction. For $\hat B $ near and larger than $ \hat B_c$,
this formula may be approximated by
\bea
\label{shatapp}
\sh \sim  {\pi \over 18}\left( {\Bh_c \over \Bh -\Bh_c}\right) \Th 
\eea
which nicely reproduces the numerical result $c_3=0.045$ obtained in formula (3.9)
of \cite{D'Hoker:2010rz} (where $k=2/\sqrt{3}$), after taking into account the change in the normalization 
of $\hat s$ and $\hat T$ between this paper and \cite{D'Hoker:2010rz}. The exact 
correspondence is 
\bea
c_3 = {\pi \hat B_c^2 \over 18 (1+\hat B_c^3)^{1/3}}
\eea 
giving approximately  $c_3=0.04186$.

 \subsubsection{Low temperature thermodynamics at $\Bh=\Bh_c$}

Next, we examine the low temperature entropy at fixed $\Bh =\Bh_c$, 
which means $\tilde \a=0$, or
\bea
\alpha= 16 c_V^2 c_{E0}^2 J(k)= {4\over 3} c_V^3 c_{E0}^2  \Bh_c^3 
\eea
We again set  $\eps_B=0$ in order to hold $c_{E0}$ fixed, and thus (\ref{epsn}) holds.   From 
(\ref{asmatch}) we also have $\eps_T=m$, and therefore 
\bea
C_T m =n
\eea
We then find
\bea
\label{Tthird}
\sh = a \Th^{1/3} 
\eea
with the numerical coefficient given by 
\bea
\label{acoeff}
 a = {\left(2\pi C_T c_V^5 c_{E0}^4 J(k)^2\right)^{1/3} \over 3}
\eea
Since $C_T c_{E0}^4$ is a universal number, which depends only on $k$ and on the properties of the 
$T=0$ purely magnetic solution, so does the  coefficient $a$. In particular, it is independent of
temperature, magnetic field and charge density.

\subsubsection{Scaling function}
\label{scaling}

Turning on both $\eps_T$ and $\eps_B$ allows us to explore a two-dimensional region around
the critical point, corresponding to changing the temperature and magnetic field.  Using
the matching relations, the change in magnetic field  to first order in $\eps_B$ can be expressed as
\bea
{\Bh -\Bh_c\over \Bh_c}  =
-  {2\over 3} {C_B \over c_{E0}} \eps_B={2\over 3} {C_B \over c_{E0}} (C_T m -n)
\eea
Combining this with 
\bea
 \Th = {1\over 4\pi}{m\sqrt{c_V n }\over \sqrt{\alpha}} 
\eea
lets us write the following cubic equation for $n$
\bea
n^3 + {3c_{E0} \over C_B} {\Bh -\Bh_c\over \Bh_c} n^2 
+ \left(  {3c_{E0} \over  2 C_B} {\Bh -\Bh_c\over \Bh_c}\right)^2 n 
- {(4\pi)^2 \alpha C_T^2 \over c_V}\Th^2 =0 
\eea
Here $\alpha $ takes its value at the critical point, $\alpha = 16 c_V^2 c_{E0}^2 J(k)$. In (\ref{CB}) of section 5, we will obtain the following result for $C_B$
\bea
C_B = 16 k J(k)^2 c_V^2 c_{E0}^3 = { 4 \over 3} k c_V^3 c_{E0}^3 J(k) \hat B_c^3
\eea
The entropy density is given by $\sh = \sqrt{c_V n \alpha}/24$, as in (\ref{stforms}). 
Combining these facts, together with the result (\ref{acoeff}), we find
that the entropy density can be expressed in the form 
\bea
\sh = \Th^{1/3} f\left( {\Bh - \Bh_c\over \Th^{2/3}}\right) 
\eea
with the scaling function $f$ obeying 
\bea
f(x) \left ( f(x)^2 + { x \over 32k\hat B_c^4} \right ) = a^3
\eea

 For $\hat B > \hat B_c$, and 
$T$ small, we have  $x>0$ and large, so that $f$ may be approximated by,
\bea
f(x) \sim { 32k \hat B_c^4 a^3 \over x}\quad \Rightarrow \quad 
{ \hat s \over \hat T} \sim {32k \hat B_c^4 a^3 \over \hat B - \hat B_c}
\eea
As we have already obtained an exact expression for the coefficient of $1/(\hat B - \hat B_c)$
for the ratio $\hat s/\hat T$ in (\ref{shatapp}),  we readily derive an expression for $a$, and
thus for $C_T$ in terms of the other parameters,
\bea 
 a^3  = { \pi \over 576 k\hat B_c ^3}~,\quad\quad 
 C_T  = { 27 \over 8 k c_V^7 c_{E0}^4 \hat B_c^9}
\eea
We have already established earlier that these values are in perfect agreement with
our numerical results of \cite{D'Hoker:2010rz} in this regime. For $\hat B= \hat B_c$, 
we have $x=0$, $f(0)=a$, and we recover the critical scaling law of (\ref{Tthird}). 

\subsubsection{Dynamical critical exponent}

The change in power law in the $\sh$ versus $\Th$ relation at the critical point is a reflection
of the change in the dynamical critical exponent.   This can made precise as follows.  As we lower
the temperature to zero, the near horizon solution is becoming BTZ $\times R^2$, with 
parameters $m$ and $n$.   Changing the temperature in this regime corresponds to a scale 
transformation of the BTZ solution, under which $m$ and $n$ transform.  Under a scale transformation
(\ref{scale}) we have
\bea
m \rt \lambda^{1-2w}m~,\quad n\rt \lambda^{2w-2} n
\eea 
For $\Bh> \Bh_c$,  taking the temperature
to zero at fixed $\Bh$ corresponds to taking $n\rt 0$ at fixed $m$.  Thus we have $w=1/2$, and
then from (\ref{zw}) this yields $z= 1$.  On the other hand, at the critical point, $\Bh=\Bh_c$,
we saw that we should take both $m$ and $n$ to zero with $m/n$ fixed.   This requires that
we take $w=3/4$, and so $z=3$. 

\subsubsection{Comparison with the numerical results on scaling of \cite{D'Hoker:2010rz}}

The value of $\hat s /\hat T^{1/3}$ at $\hat B = \hat B_c$ was 
encoded in the parameter $c_2$ of \cite{D'Hoker:2010rz}, where $k$ was fixed at the 
supersymmetric value $k=2/\sqrt{3}$.  The precise relation,
taking into account the required conversion of normalizations of $\hat s$ and 
$\hat T$ between this paper and \cite{D'Hoker:2010rz},  is given by 
\bea
c_2 = { a \hat B_c^{4/3} \over (1+\hat B_c^3)^{4/9}}
\eea
We find the value $c_2=0.1265$, which compares favorably with the numerical outcome 
$c_2=0.11$ of \cite{D'Hoker:2010rz}. Finally, as $\hat B < \hat B_c$, we find 
\bea
f(x) \sim {(-x)^{1/2} \over 4 \sqrt{2k} \, \hat B_c^2 } \hskip 1in 
\hat s \sim { \sqrt{\hat B _c - \hat B} \over 4 \sqrt{2k} \, \hat B_c^2} 
\eea
This result may be compared with the coefficient $c_1$ of formula (3.7) in 
\cite{D'Hoker:2010rz}.
Taking into account the conversion of conventions, we find 
\bea
c_1 = {1 \over 4 \sqrt{2k} (\hat B_c)^{1/2} (1+\hat B_c^3)^{1/2}}
\eea
which gives $c_1 = 0.221$ as compared with the 
numerical value $c_1=0.172$ of \cite{D'Hoker:2010rz}.

\subsubsection{Approach to zero temperature in the $(b,q)$ plane}
\label{bqcrit}

The data used to parametrize the solutions studied numerically
in \cite{D'Hoker:2009bc} and \cite{D'Hoker:2010rz} consisted of the charge density
at the horizon $q$, as well as the magnetic field strength $b$ in the horizon frame.\footnote{The 
normalization of the parameter $q$ used in \cite{D'Hoker:2010rz} differs from the 
normalization of the parameter $q$ used here by a factor of $\sqrt{n}$, so that 
$\sqrt{n} q_{here} = q_{\cite{D'Hoker:2010rz}}$. In this subsection only, $q$ will denote the
parameter of \cite{D'Hoker:2010rz}.}
(A third parameter, denoted by $C'(r_+)$, was associated with the momentum $P_3$
of the solution, and could be fixed arbitrarily.) In this parametrization, the electrically
charged, finite temperature,  Reissner-Nordstrom solutions correspond to $b=0$, 
while the purely magnetic finite temperature solutions of \cite{D'Hoker:2009mm}
correspond to $q=0$, their extremal $T=0$ limits being reached respectively at 
the endpoints $(b,q)=(0,\sqrt{6})$ and $(b,q)=(\sqrt{3},0)$.    

\sm

For $\Bh \geq \Bh_c$  (and for $k\geq 3/4$, as will become clear below), 
lowering the temperature to zero corresponds in terms of $(b,q)$ to approaching
the purely magnetic endpoint $(b,q)=(\sqrt{3},0)$.      This circumstance
further illuminates the remark made in the last paragraph of section \ref{critmag}
that the low $T$ thermodynamics, for all values of $\hat B \geq \hat B_c$, are governed by 
solutions which are all related to one another by coordinate transformations 
at $T=0$.  We emphasize that even at
this endpoint the full asymptotically AdS$_5$ solution carries a nonzero charge density; what is
tending to zero here is the charge at the horizon. 

\sm

The departure infinitesimally away from the {\sl purely magnetic fixed point} at $(b,q)=(\sqrt{3},0)$
caused by turning on a small $q$ is known analytically from our work here.
To see this, note that the value $v_0$ of the field $V$ at the horizon is equivalent, via rescaling
of $x_{1,2}$, to a
change in $b$, which is given to leading order by, $\delta b  = \sqrt{3} -b= - 2 v_0$.
With the present definition of $q$, the matching equations obtained in  (\ref{match7}) read
\bea
\label{match7mod}
 {E_k q /\sqrt{n}\over (mn)^{k-1}} = e_0~,\quad\quad
   {V_\sigma  \over (mn)^\sigma} (v_0 + q^2 A_q)=1 
\eea
The parameter $e_0$ is kept fixed, since it is related to the value of $\Bh$,  which is being
held fixed as we lower the temperature.  
Rearranging and trading $v_0$ for $\delta b$, we have
\bea
\label{matchb}
 q = {e_0 \over E_k} m^{k-1}n^{k-{1\over 2}} ~,\quad\quad  \delta b = 2A_q  q^2 -2 (V_\sigma )^{-1} (mn)^\sigma
\eea
The parameter $A_q$, which will be obtained in section \ref{match1}, is negative when 
the $z'$-integral is convergent, namely for $k< 1+\sigma/2$. These relations allow us to 
examine the curve in the $(b,q)$-plane along which the $T=0$ limit is obtained
while keeping $\hat B$ fixed. 

\medskip

$\bullet$ For $\Bh > \Bh_c$, the zero temperature limit is obtained by taking $n$ to zero while
holding $m$ fixed.    It is then appropriate to write 
\bea
\delta b = 2A_q q^2 - 2 V_\sigma^{-1} m^{\sigma \over 2k-1} 
\left(E_k \over e_0\right)^{\sigma \over k-{1\over 2}} q^{\sigma \over k-{1\over 2}} 
\eea
The first   term dominates for ${1\over 2} < k<  {\sigma+1  \over 2}$; 
the second for  $   {\sigma+1  \over 2} <k$. In either case the $q \to 0$
limit is smooth, since both powers of $q$ are positive. Thus, as $T \to 0$, the $\hat B > \hat B_c$
system  flows to the purely magnetic critical point for all $k > 1/2$. 
Numerical study confirms these analytical results.

\medskip

$\bullet$ For $\Bh = \Bh_c$, the zero temperature limit is  obtained by taking both $m$ 
and $n$ to zero while holding the ratio $m/n$ fixed. We write
\bea
\label{deltab}
\delta b = 2A_q q^2 -2 V_\sigma^{-1} \left(E_k \over e_0\right)^{\sigma \over k-{3\over 4}}
\left( m\over n\right)^{\sigma \over 4k-3} q^{\sigma \over k-{3\over 4}} 
\eea
For $ {\sigma \over 2}+{3\over 4} <k $ the second term dominates, 
and thus determines the flow.
For ${3\over 4} < k <  {\sigma \over 2}+{3\over 4}  $ the first term dominates.  

\sm

For $  {3 \over 4}<k$, both terms admit a smooth $q \to 0$ limit since, just as in the case 
$\hat B > \hat B_c$,  both powers of $q$ are positive. Thus, as $T \to 0$, 
the $\hat B=\hat B_c$ system   again flows to the purely magnetic critical point, 
a result again confirmed by numerical study.

\sm

Finally, for  ${1\over 2} < k < {3\over 4}$ the power of $q$ in the second term turns negative, 
indicating that the system no longer flows towards the purely magnetic critical point as $T \to 0$.  
In fact, numerical study shows that as $T \to 0$, a finite limiting value for $q$
emerges. Thus, the perturbative expansion around the purely magnetic critical point,
which was used throughout this paper, can no longer be valid for the $\hat B = \hat B_c$
and ${1\over 2} < k < {3\over 4}$ system, and the scaling law $\hat s \sim \hat T^{1 \over 3}$
is not expected to hold. The precise nature of the flows for $ k < {3\over 4}$ 
is presently under investigation, and will be discussed elsewhere \cite{inprog}.

\sm

The critical curve, defined as the set of limiting values in the $(b,q)$ plane at which $T=0$,
may be computed from the above result, in the neighborhood of the purely magnetic
critical point, where it may be defined as the flow at $\hat B = \hat B_c$. We deduce from
(\ref{deltab}) that it is given by $\delta b \sim -q^2$ for $3/4< k< 3/4+\sigma/2$, and by 
$\delta b \sim - q^{\sigma /(4k-3)}$ when $k > 3/4 + \sigma/2$, a result found to be in 
accord with numerical results.

\section{Matched asymptotic expansion: detailed analysis}
\setcounter{equation}{0}
\label{matching}

To explore the low temperature thermodynamics in the regime $ B \geq B_c$, 
we proceed perturbatively in small $ T$, namely $ T \ll \sqrt{ B_c}$, 
as outlined in section \ref{LowT}. For $r \gg T$, the effects of finite temperature
are small and may be treated perturbatively; in this region, we use a perturbative 
solution around
the charged, $T=0$, asymptotically $AdS_5$ solution constructed in section~\ref{ZeroT}.
For $r \ll \sqrt{B_c}$, the finite $T$ asymptotically $AdS_5$ solution reduces to 
BTZ $ \times R^2$;
in this region, we use a perturbative solution around finite temperature BTZ. 
In the overlap region $T \ll r \ll \sqrt{B_c}$ the effects of temperature are small and a 
perturbed $AdS_3$ solution may be used. It is in this region that both the near-horizon 
perturbed BTZ solution and the asymptotically $AdS_5$ perturbed solutions are 
both valid, and where their perturbative solutions may be matched. In this section,
we present detailed derivations of the perturbative expansions in each one of these 
regions, and of their matching.  This section is somewhat technical, and so we note
that the main results have already been summarized in the preceding section. 

\subsection{Near-horizon region: perturbed BTZ solution}

In the near-horizon region, we expand around the BTZ $ \times R^2$ solution of 
(\ref{BTZ1}) and (\ref{BTZ2}). Inspection of the reduced field equations (\ref{baneqs})
instructs us to treat $E$ and $P$ to first order, but the other fields to second order.
To organize this perturbation theory we introduce a small parameter $\ep$, so that
\bea
\label{BTZpert}
E(r) & = &  \ep E_1(r)
\no \\
P(r) & = &  \ep P_1(r)
\no \\
L(r) & = & 2br + \ep^2 L_1(r)
\no \\
M(r) & = & -mr + \ep^2 M_1(r)
\no \\
N(r) & = & n + \ep^2 N_1(r)
\no \\
V(r) & = &  \ep^2 V_1(r)
\eea
where $b=\sqrt{3}$, and $m$ and $n$ are the constant parameters of the BTZ solution.
The corresponding field equations are,
\bea
\label{BTZeq2}
M1 \qquad 0 & = &  n E_1' + 2 br P_1' +2b(k+1)P_1
\\
M2 \qquad  0 & = & 2b r E_1' +2b(1-k)E_1-mrP_1'-mP_1
\no \\
E1 \qquad  0 & = & \left ( L_1' + 8 b r V_1' - 4 bV_1 \right )' - 4 E_1P_1
\no \\
E2 \qquad 0 & = & \left ( M_1' - 4 mr V_1' + 2 m V_1 \right ) ' + 4 E_1^2
\no \\
E3 \qquad 0 & = & \left ( N_1' + 4 n V_1' \right )' + 4 P_1^2
\no \\
E4' \qquad 0 & = & 6 \Big ( ( 12 r^2 + mnr) V_1' \Big )'  - 96 V_1 
+ 4 n E_1^2 + 16 br E_1 P_1 - 4 mrP_1^2
\no \\
CON \qquad 0 & = & b L_1' + { m \over 4} N_1 ' +(24r+mn)V_1' -12 V_1
+ nE_1^2+4brE_1P_1-mrP_1^2
\no
\eea
We have replaced E4 by E4'= E4 + 4 CON. The boundary conditions at the 
horizon are, 
\bea
\label{bc1}
L_1(0)=M_1(0)=M_1'(0)=N_1(0)=0 
\eea
The first two may be chosen by SL(2,R) and the requirement that the horizon remain at $r=0$; 
the latter two follow from the fact that any non-zero values may be absorbed into
the parameters $m,n$ of the zero-th order solution. The values of $E_1(0)$ and $V_1(0)$
will be turned on by the perturbation, and $P_1(0)$ will be determined by those, as will
be shown below.

\subsubsection{Solving Maxwell's equations}

Maxwell's equations M1 and M2 involve only $E_1$ and $P_1$ and may thus be 
integrated independently from the remaining equations. Throughout, it will often be 
convenient to use the rescaled coordinate $z$ defined by,
\bea
z \equiv - { 12 r \over mn}
\eea
We shall use the same notation for a function of $r$ and its associated function of $z$.
Eliminating $P_1'$ between M1 and M2, it follows that $P_1$ is uniquely determined by $E_1$,
\bea
P_1 = {\sqrt{12} \over  km} \bigg [ (1-z) E_1' +  (k-1) E_1 \bigg ]
\eea
while $E_1$ satisfies the hypergeometric differential equation,
\bea
\label{hyper}
z(1-z) \p_z^2 E_1 + [c-(a+b+1)z] \p_z E_1 - ab E_1=0
\eea
with $a+b=2$, $ab=1-k^2$, and $c=1$. Retaining the solution which is regular at $r=0$, 
we impose a boundary condition $E_1(0)=q$, corresponding to finite charge density at the horizon. 
The solution is then given by the hypergeometric function,
\bea
E_1(r) = q F(1+k, 1-k; 1; z)
\eea
The functions $E_1$ and $P_1$ are both regular as $r \to 0$, taking values 
\bea
E_1(0)=q \hskip 1in P_1(0)= -\sqrt{12} q(k-1)/m
\eea

\subsubsection{Solving Einstein's equations}

Einstein equation E4' may be solved for $V_1$ in terms of $E_1$ and $P_1$.
To do so, it will be useful to introduce the composite function $\Lambda$, defined by,
\bea
n q^2 \Lambda \equiv { 1 \over 18} \left (n E_1^2 + 4 b r E_1 P_1 - mr P_1^2 \right )
\eea
By inspection, it is clear that the function $\Lambda$ depends only on $z$ and $k$. In terms of these
variables, E4' for $V_1$ may be expressed as an inhomogeneous hypergeometric equation,  
\bea
z(1-z) \p_z^2 V_1 +(1-2z) \p_z V_1 + { 4 \over 3} V_1 = n q^2 \Lambda (z)
\eea
The homogeneous part is the hypergeometric equation with $a+b=1$, $ab=-3/4$, and $c=1$,
following the notation of (\ref{hyper}).
As a result, $a$ and $b$ are given by $a = 1+\sigma$ and $b = -\sigma$, where 
$\sigma$ was defined in (\ref{sigma}). There is a unique solution which is regular at $r=z=0$,  which we shall abbreviate by,
\bea
F(z) \equiv F(\sigma +1, - \sigma; 1; z)
\eea
To construct the inhomogeneous solution, we set $V_1(z) = F(z) v(z)$, after which 
$\p_z v$ is found to satisfy a first order equation, 
\bea
\p_z \left ( z(1-z) F^2 \p_z v \right ) =  n q^2 F \Lambda
\eea
This equation may be readily integrated, and the resulting general solution which is 
regular as $ z= r = 0$ is given by,
\bea
\label{V1full}
V_1  (z) = v_0 F(z)
+ n q^2 F(z) \int _0 ^z dz' { 1 \over z'(1-z') F(z')^2 } \int _0 ^{z'} dz'' F(z'') \Lambda (z'')
\eea
Here, $v_0$ is an arbitrary integration constant. Since $E_1$ and 
$P_1$ are regular as $z \to 0$, both integrals are convergent, so that $V_1(z)$
is a well-defined function of $z$ with $V_1(0)=v_0$.

\sm

Einstein's equations E1, E2, E3 may be solved in terms of $E_1, P_1, V_1$, as follows, 
\bea
L_1(r)  & = & 
4 \a_Lr - 8 b r V_1(r) + 12 b\int _0 ^r dr' V_1(r') + 4 \int _0 ^r \! dr' \int ^{r'} _0 \! dr'' E_1P_1(r'')  
\no \\
M_1(r) & = & 
2 v_0 r  + 4 m r V_1(r) - 6 m \int _0 ^r dr' V_1(r')  - 4 \int_0 ^r \! dr' \int ^{r'} _0 \! dr'' E_1^2(r'') 
\no \\
N_1(r)  & = &  4 \a_Nr  - 4 n (V_1(r) -v_0) - 4 \int _0 ^r \! dr' \int ^{r'} _0 \! dr'' P_1^2(r'') 
\eea
These solutions fulfill all the boundary conditions of (\ref{bc1}). The two remaining integration 
constants $\a_L,\a_N$ are related by enforcing the constraint equation CON, and we find, 
\bea
n q^2 = - 4 b \a_L - m \a_N
\eea
In summary, given $m,n$ of the unperturbed BTZ solution, the perturbation theory 
around BTZ
is governed entirely by two additional parameters $q$, $v_0$ (the parameter $\a_L$
will be determined by further gauge fixing and matching).

\subsection{Overlap region: perturbed $AdS_3$ solution}

In the overlap region, the unperturbed geometry is simply $AdS_3 \times R^2$, whose fields
are given in (\ref{AdS3}) with $V=0$, and corresponds to setting $m=n=0$ in the BTZ equations 
(\ref{BTZpert}) and (\ref{BTZeq2}). All fields are now expanded to first order only, and we
find the leading order solutions,
\bea
\label{overlap}
E(r) & = & e_0 r^{k-1}
\no \\
P(r) & = & p_0 r^{-k-1}
\no \\ 
L(r) & = & l_0 + 2 b r -4\sqrt{3}\left(2\sigma -1 \over \sigma+1\right) v_+ r^{\sigma+1}
 -4\sqrt{3}\left(2\sigma +3 \over \sigma\right) v_- r^{-\sigma}
\no \\
M(r) & = & m_0 + m_1 r
\no \\
N(r) & = & n_0 + n_1 r
\no \\
V (r) & = & v_+ r^\sigma + v_- r^{-\sigma -1}
\eea 
Note that the constraint equation CON fixes the linear term in $L$.
There are 9 integration constants, $ e_0, p_0 , l_0 ,  m_0 , m_1,  \tilde n_0 , n_1, v_+ , v_-$; 
their number precisely matches the expected number from 4 second order equations, 
2 first order equations and one constraint.

\subsection{Matching expansions in near-horizon and overlap regions}
\label{match1}

To carry out the matching between these two regions, we need to isolate, 
in the large $r/(mn)$ behavior of the perturbative solution around the BTZ solution,
those functional dependences that coincide with those identified in the 
overlap region in (\ref{overlap}). These large $r/(mn)$ asymptotics arise 
from the asymptotic behavior for large $z$ of the hypergeometric function,
which may be obtained using the inversion formula, 
\bea
\label{hyper1}
F(a;b;c;z)  &=& 
{\Gamma(c)\Gamma(b-a)\over \Gamma(b)\Gamma(c-a)} (-z)^{-a}
F\left(a;1-c+a;1-b+a;z^{-1}\right) 
\\ \no 
& +&  
{\Gamma(c)\Gamma(a-b)\over \Gamma(a)\Gamma(c-b)} (-z)^{-b}
F\left(b;1-c+b;1-b+a;z^{-1} \right) 
\eea
which results in the following dominant asymptotics as $-z \to \infty$ ,
\bea
\label{hyper2}
F(a;b;c;z)  \sim  
{\Gamma(c)\Gamma(b-a)\over \Gamma(b)\Gamma(c-a)} (-z)^{-a}
+{\Gamma(c)\Gamma(a-b)\over \Gamma(a)\Gamma(c-b)} (-z)^{-b}
\eea
Isolating the $r^{k-1}$ term in $E_1$ and the $r^{-k-1}$ term in $P_1$, we find the following
relations, 
\bea
e_0 & = & q (mn)^{1-k} E_k  \hskip 0.8in E_k \equiv {(12)^{k-1} \G (2k) \over \G (k) \G (k+1)}
\no \\
p_0 & = &  q (mn)^{1+k} E_{-k} 
\eea
Isolating the $r^\sigma$ and $r^{-1-\sigma}$ terms in $V_1$ proceeds analogously,
paying close attention to some extra subtleties. We begin with the asymptotics
as $r/(mn) \to \infty$ of the corresponding hypergeometric function, 
\bea
F(z) \sim  V_\sigma (mn)^{-\sigma} r^\sigma +  V_{-1-\sigma} (mn) ^{1+\sigma} r^{-1-\sigma}
\hskip 0.8in 
V_\sigma \equiv { (12)^\sigma \G(1+2\sigma) \over \G(1+\sigma)^2}
\eea
The subtlety in evaluating the asymptotics of $V_1$ resides in the fact that 
contributions arise due to the inhomogeneous part of the solution in (\ref{V1full}).
To identify these contributions, one needs to investigate the asymptotics of 
the integrals for large $-z$. The behavior of $\Lambda$ is readily deduced 
from that of $E_1$ (given in (\ref{overlap})), and that of $P_1$, given by
\bea
P_1(z) \sim - { 4 b q (k-1) E_{k-1} \over mk (mn)^{k-2} }  r^{k-2}
\eea
As a result, the $-z\to \infty$ asymptotics of $\Lambda$ is given by,
\bea
\Lambda (z) \sim \Lambda _\infty (mn)^{2-2k}  r^{2k-2}
\hskip 1in \Lambda _\infty = { E_k^2 \over 18 (2k-1)} 
\eea
The asymptotics of the  integral over $z''$ in (\ref{V1full}) as $-z \to \infty$
is proportional to $(-z')^{2k-1+\sigma}$. For $k>1/2$, as we have been assuming
throughout, this exponent is positive. The leading asymptotics of the $z'$ 
integral in (\ref{V1full}) is then governed by the asymptotics of the integrand,
namely $(-z')^{2k-3-\sigma}$. For $2k-2<\sigma$, the integral converges as
$z \to \infty$, and the leading asymptotics of $V_1$ is given by
\bea
\label{AVV1}
V_1(r) \sim A_V V_\sigma (mn)^{-\sigma} r^\sigma +  
A_V V_{-1-\sigma} (mn) ^{1+\sigma} r^{-1-\sigma}
\eea
where $V_\sigma$ was defined earlier, and $A_V$ is given by,
\bea
\label{AV}
A_V = v_0 + n q^2 A_q
\eea
where $A_q$ is a quantity that depends only on $k$. For $2k-2<\sigma$
(or equivalently $ k < 1+\sigma /2 \sim 1.38$, so that this range clearly includes the supersymmetric value of $k$), we have 
\bea
\label{AV1}
A_q \equiv \int _0 ^{-\infty} dz' { 1 \over z'(1-z') F(z')^2 } \int _0 ^{z'} dz'' F(z'') \Lambda (z'')
\eea
For larger values of $k > 1+\sigma /2$, there is still a contribution of the form
(\ref{AV}), but it must now be obtained after subtracting out the leading $r \to \infty$
asymptotics of the integral.  For example, when $\sigma < 2k-2< \sigma +1$, we have
\bea
A_q= \lim _{-z \to \infty} \left [
\int _0 ^z dz' { 1 \over z'(1-z') F(z')^2 } \int _0 ^{z'} dz'' F(z'') \Lambda (z'') 
- { v_\infty (-z)^{2k-2} \over F (z)} \right ]
\eea
where 
\bea
v_\infty = - { 3 \Lambda _\infty \over 2 (6k^2-9k+1)}
\eea
Every increment in the range of $k$ by 1/2 will require an extra  subtraction term.

\sm

We are now ready to match the asymptotics of $V_1$ between the near-horizon 
and overlap regions, and we find,
\bea
\label{AVV2}
v_+ & = & A_V V_\sigma (mn)^{-\sigma}
\no \\
v_- & = & A_V V_{-1-\sigma} (mn)^{1+\sigma}
\eea
The ratio $v_- /v_+ = (V_{-1-\sigma}/V_\sigma) (mn)^{1+2 \sigma}$
shows that the term proportional to $v_-$ corresponds to a higher order 
correction in the asymptotically $AdS_5$ region, and may be neglected there, 
so that we effectively have $v_- \sim 0$.

\sm

It remains to match the expansions of the functions $L_1, M_1, N_1$ . To leading order, 
we have the matching, 
\bea
\label{LMNmatch}
l_0 = 0 & \hskip 1in &
\no \\
m_0 =0 && m_1 = -m
\no \\
n_0 = n && n_1 = 0
\eea
The large $r/(mn)$ asymptotics of the functions $L_1, M_1, N_1$ 
may be derived in an analogous fashion, and will provide higher order corrections 
to the leading matching of (\ref{LMNmatch}). The corresponding results will not,
however, be needed for the thermodynamic questions that we are addressing
here, and we shall not carry them out.

\subsection{Asymptotically $AdS_5$ region: perturbed $T=0$ charged solution}

In the asymptotically $AdS_5$ region, perturbation theory is carried out around 
the $T=0$ charged asymptotically $AdS_5$ solution, which was derived in
section \ref{ZeroT}. Here, we shall denote the fields of this solution with 
subscripts 0. The functions $L_0$ and $V_0$ obey the equations,
\bea
\label{LVeq}
0 & = & L_0'' + 2 V_0' L_0' + 4 L_0 (V_0'' +(V_0')^2) 
\no \\
0 & = &  L_0^2 (V_0')^2 + V_0' (L_0^2) '  -6 +{1\over 4}(L_0')^2  + 3 e^{-4V_0}
\eea
for both neutral and charged solutions. For the charged 
solutions, we have $E_0=A_0'$ with,
\bea
\label{Azero1}
A_0(r) & = & {c_V c_{E0} \over kb} e^{2kb \psi(r)} 
\no \\
M_0(r) & = & - {\a \over \sqrt{12}}  L_0(r)  
- { 4 c_V^2 c_{E0}^2 \over kb} L_0(r) 
\int ^r _\infty dr' { e^{4kb \psi (r')} \over L_0(r')^2 e^{2V_0(r')}}
\eea
where the function $\psi$ is defined by
\bea
\hskip 1in 
\psi (r) \equiv \int ^r _\infty { dr' \over L_0(r') e^{2V_0(r')}}
\eea
as well as $N_0=P_0=0$. For given $V_0$, the first equation of (\ref{LVeq}) 
is a linear second order differential equation. Besides it solution $L_0$,
it has a conjugate linearly independent solution, which we shall denote  
by $L_0^c$, and normalize by,
\bea
\label{Lceq}
L_0^c (r) \equiv L_0(r) \int _\infty ^r { dr' \over L_0(r')^2 e^{2V_0(r')}}
\eea
This function obeys the following asymptotics,
\bea
r \to 0 & \hskip 1in & L_0 ^c (r) \to - { 1 \over \sqrt{12}}
\no \\
r \to \infty && L_0^c(r) \sim -{ 1 \over 4c_V r}
\eea
This solution will play a key role in the sequel. 

\subsubsection{Perturbation equations}

To carry out first order perturbation theory around this solution, we introduce 
an expansion parameter $\ep$, and set, 
\bea
\label{expeps}
P(r) & = &  \ep P_1(r)
\no \\
N(r) & = &  \ep N_1(r) 
\no \\
E(r) & = & E_0(r) + \ep E_1(r)
\no \\
L(r) & = & L_0(r) + \ep L_1(r)
\no \\
M(r) & = & M_0(r) + \ep M_1(r)
\no \\
V(r) & = & V_0(r) + \ep V_1(r)
\no \\
f(r) & = & f_0 (r) + \ep f_1 (r) 
\eea
Maxwell's equations for the perturbation functions are given by,
\bea
\label{pertmax}
M1 & \hskip 0.4in & 0 = \left ( e^{2V_0} ( E_0 N_1 + L_0 P_1) \right ) ' + 2 k b P_1
\no \\
M2 && 0 = \left ( e^{2V_0} ( 2L_0 E_0 V_1 + L_0 E_1 +E_0 L_1 +M_0 P_1) \right )' - 2 kb E_1 
\eea
while Einstein's equations are given by, 
\bea
\label{pertein}
E1 & \hskip 0.3in & 
0 = L_1'' + 2 V_0 ' L_1' + 2 V_1' L_0' + 4 L_1 \left ( V_0'' +(V_0')^2 \right ) 
+ 4 L_0 \left ( V_1'' + 2 V_0' V_1'  \right ) - 4 E_0 P_1 
\no \\
E2 && 0 = M_1'' + 2 V_0 ' M_1' + 2 V_1' M_0' + 4 M_1 \left ( V_0'' +(V_0')^2 \right ) 
+ 4 M_0 \Big ( V_1'' + 2 V_0' V_1' \Big ) + 8 E_0 E_1 
\no \\
E3 && 0 = N_1'' + 2 V_0 ' N_1' + 4 N_1 \left ( V_0'' +(V_0')^2 \right ) 
\no \\
fV && 0 = f_1'' + 4 V_0' f_1' + 4 V_1' f_0' + 2 V_1'' f_0 
+ 2 V_0'' f_1 + 4 (V_0')^2 f_1 + 8 V_0' V_1' f_0
\no \\
E4' && 0 = \left (6V_0'' + 12 (V_0')^2 \right ) f_1 
+ \Big (6 V_1'' + 24 V_0 ' V_1' \Big ) f_0  + 6 V_1' f_0' + 6 V_0' f_1'
\no \\ && \qquad
 - 32b^2 e^{-4V_0} V_1 + 8 L_0 E_0 P_1 + 4 N_1 E_0^2
\eea
Here, the last equation corresponds to the combination $E4'=E4+4CON$,
while the next-to-last equation derives from expanding the $f,V$ equation of (\ref{fV}).

\subsubsection{Boundary conditions}

The $T\not=0$ perturbations on the $T=0$ asymptotically $AdS_5$  solutions 
will be required to have specific boundary conditions at $r=\infty$, namely
that the space-time retain the same asymptotic metric as the $T=0$
solution. This will require $L_1(r)/r$, $M_1(r)/r$, $N_1(r)/r$, $r^2 E_1(r)$,
and $V_1(r)$ to tend to 0 as $r \to \infty$. At small $r$, namely $T\ll r \ll \hat B_c$,
we will impose matching conditions with the perturbations around the 
$T\not=0$ BTZ solution, as given in the overlap region by (\ref{overlap}).

\subsubsection{The translation and dilation modes}

Invariance of the equations (\ref{baneqs}) for the $T=0$ unperturbed asymptotically 
$AdS_5$ solution under translations and dilations in $r$ guarantee the existence of two 
independent perturbative solutions to (\ref{pertmax}) and (\ref{pertein}).
They will be denoted with the superscripts $t$ and $d$ respectively,
and are given by,
\bea
\label{tdeq}
E_1^t =E_0' ~& \hskip 1in & ~E_1 ^d =r E_0' + E_0/2
\no \\
L_1^t =L_0' ~& \hskip 1in & ~L_1 ^d =r L_0' -L_0
\no \\
M_1^t =M_0' & \hskip 1in & M_1 ^d =r M_0' - M_0
\no \\
V_1^t =V_0' ~& \hskip 1in & ~V_1 ^d =r V_0' 
\no \\
f_1^t =f_0' ~& \hskip 1in & ~f_1 ^d =r f_0' -2 f_0
\eea
with $P_1^t = P_1 ^d =N_1^t = N_1 ^d =0$. As expected, the translation mode 
obeys the desired boundary conditions at $r \to \infty$, but the dilation mode 
does not, because $V_1^d(r)$ and $r^2 E_1^d(r)$  do not tend to 0 there. 
Thus, the  dilation mode must be absent altogether.

\subsubsection{Solving and matching for $N_1$}

The function $N_1$ satisfies the same differential equation as $L_0$ does.
As a result, it must be a linear combination of $L_0$ and its conjugate $L_0^c$.
The boundary conditions at $r\to \infty$ forces the function $L_0$ to be absent, 
so that we have,
\bea
\label{N1eq}
N_1 (r) = \tilde n_0 L_0^c (r) 
\eea 
with asymptotics given by,
\bea
\label{N1as}
r \to 0 & \hskip 0.5in & N_1 (r) \sim - {\tilde n_0 \over \sqrt{12}} 
\no \\
r \to \infty & \hskip 0.5in & N_1 (r) \sim   - { \tilde n_0 \over 4 c_V r}
\eea
Matching the $r\to 0$ asymptotics with those of the overlap region, we find that
\bea
n_0 = - { \tilde n_0 \over \sqrt{12}} \hskip 1in n_1=0
\eea
since the function $L_0^c$ does not contain a dependence on $r$
which is linear for small $r$.

\subsubsection{Solving and matching for $P_1$}

The function $P_1=C_1'$ may be obtained by solving M1 in (\ref{pertmax}).
It is instructive, however, to solve for $C_1$ from the first integral 
for $\lambda$ in (\ref{einstint}). Linearizing this equation gives,
\bea
(N_1 M_0' - N_1' M_0) e^{2 V_0} = 2 \lambda _0 + 8 kb A_0 C_1
\eea
This equation is easily solved, and we find the general solution, 
\bea
C_1 = - \tilde n_0 { A_0  L_0 ^c \over 2 L_0}  
- { \tilde n_0 M_0 \over 8kb L_0A_0} - { \lambda _0 \over 4kb A_0}
\eea
The relevant asymptotics of $P_1$ as $r \to 0$ is the functional dependence $r^{-1-k}$,
resulting from the dependence $r^{-k}$ of $C_1$. The first term above behaves
as $r^{k-1}$ which has no overlap with $r^{-1-k}$ for $k >1/2$. The last two
terms do contribute, however, and we find, 
\bea
\label{C1as}
 r \to 0 & \hskip 0.6in & C_1 \sim { \tilde n_0 \tilde \a - 4 b \lambda _0 \over 48 e_0} r^{-k}
\no \\
r \to \infty & & C_1 \sim { \tilde n_0 \a - 4 b \lambda _0 \over 16 b c_V c_{E0}}
\left ( 1 +{ kb \over c_V r} \right )
\eea
As a result, we find,
\bea
\label{P1as}
 r \to 0 & \hskip 0.6in & P_1 \sim p_0 r^{-k-1}
 \hskip 0.74in p_0 =  - k { \tilde n_0 \tilde \a - 4 b \lambda _0 \over 48 e_0}
\no \\
r \to \infty & & P_1 \sim  {p_\infty  \over r^2} 
\hskip 1in 
p_\infty =  - k { \tilde n_0 \a - 4 b \lambda _0 \over 16  c_V^2 c_{E0}}
\eea
Comparison with the functional behavior in the overlap region, we find
that the prefactors $p_0$ must be the same in both expressions.
The value of $p_0$ in the overlap region has already been determined, 
and found to be given by $p_0 = q E_{-k} (mn)^{1+k}$. This value provides
a higher order correction and vanishes to leading order, $p_0 \sim 0$.
Combining with the result from (\ref{P1as}), we must have to this order,
\bea
\lambda _0 = \tilde n_0 \tilde \a /(4 b)
\eea
so that 
\bea
p_\infty = - \tilde n_0 k { \a - \tilde \a \over 16 c_V^2 c_{E0}} = - \tilde n_0 k c_{E0} J(k)
\eea
The large $r$ asymptotics of $P_1$ will play a key role in determining the constant $C_B$.

\subsection{Calculation of the coefficient $C_B$}

With the matching information on $N_1$ and $P_1$ in hand so far, it is 
already possible to compute
the coefficient $C_B$ which plays a key role in determining the scaling properties
near the quantum critical point. Since $C_B$ governs the response of the 
charge density $c_E$ to a change in the parameter $n$ at the horizon, we need to 
determine the change in  the field $E_1$ caused by turning on the perturbations 
$N_1$ and $P_1$. We begin by solving for $E_1$.

\subsubsection{Solving for $E_1$}

The function $E_1$ may be obtained by solving equation M2 of (\ref{pertmax}), 
in terms of the functions  $P_1,V_1, L_1$. In fact, it is instructive to obtain the 
corresponding potential $A_1$ from the first integral 
in the second equation of (\ref{maxint}). Its perturbative expansion is given by,
\bea
L_0 e^{2V_0} A_1' - 2 kb A_1 = - e^{2V_0} \Big ( 2 L_0 E_0 V_1 
+ E_0 L_1 + M_0 P_1 \Big )
\eea
Dividing both sides by $A_0 L_0 e^{2V_0}$ and using (\ref{Azero}), we readily
integrate this equation, and the general solution is given by,
\bea
\label{A1int}
A_1(r)  = A_0(r) \left [ a_1  -  \int _\infty ^r dr' 
{1 \over A_0(r')}  \left ( 2  E_0 V_1 
+ {E_0 \over L_0} L_1 + {M_0 \over L_0} P_1 \right )(r') \right ]
\eea
where $a_1$ is an integration constant which may be absorbed into the 
zero-th order charge $e_0$ or $c_{E0}$. The matching for small $r$
is carried out between the zero-th order asymptotically $AdS_5$ 
solution $E_0$ and $E(r)=e_0 r^{k-1}$ in the overlap region (\ref{overlap}),
so that $c_{E0}$ and $e_0$ are related by the zero-th order formula
(\ref{ezero}). The corrections provided by $E_1$ only add higher order effects,
which may be ignored for our purposes.

\sm

To compute the change in charge density $c_E$, we need to obtain only the 
leading correction to the $1/r$ asymptotics of $A_1$. Remarkably, to do so,
we do not need to know the precise solutions for the functions $L_1$ and $V_1$.
The only property of $L_1$ and $V_1$ that we will need here is the fact that they
preserve the asymptotic form of the metric, namely as $ r \to \infty$, the 
functions $r V_1$ and $L_1$ must have finite limits. 

\sm

As a result of the above properties of $V_1 $ and $L_1$, the leading asymptotics of the integrand of (\ref{A1int}) derives entirely from the term in $P_1$, and is of order $1/r^2$.
The integral converges at $ \infty$, and the asymptotics there is given by,
$A_1 (r)  \sim -  \a  p_\infty /( \sqrt{12} \, r)$,  where $p_\infty$ was defined in (\ref{P1as}).
As a result, the asymptotics of $E_1$ is given by
\bea
\label{E1as}
E_1(r)  \sim  { c_{E1} \over r^2} 
\hskip 1in
c_{E1} =   {\a  p_\infty   \over \sqrt{12} } 
= 16 n_0 c_V^2 c_{E0}^3 k J(k)^2
\eea
Thus,  the change in the charge density caused by turning on $n_0$ is
indeed non-trivial.

\subsubsection{Calculating $C_B$ at the critical point}

The coefficient $C_B$ is defined as the change in $c_E$ due to turning on $N$ at the horizon, 
\bea
c_E & = & c_{E0} + C_B \epsilon_B
\no \\
N & = & \epsilon_B \hskip 1in r \to 0
\eea
while keeping the temperature fixed at $T=0$. Thus, we need to know the  
asymptotics of the field $E_1$ due to turning on $n_0$. This is precisely
the effect calculated above, and those results may now be applied.
The quantity $c_{E} - c_{E0}$ may be identified with $c_{E1}$ above,
and $\epsilon_B$ with $n \sim n_0$ to leading order, so that we find,
\bea 
\label{CB}
C_B = {c_{E1} \over n_0} =   k J(k)  \a  c_{E0} = 16 k J(k)^2 c_V^2 c_{E0}^3
\eea
using the results of (\ref{E1as}). As was shown in section \ref{scaling}, once $C_B$
is known, the remaining coefficient $C_T$ may be calculated indirectly by matching
the scaling function near the quantum critical point onto the low $T$ behavior
of the $\hat B > \hat B_c$ branch, which is known exactly. In the remainder of 
this section, however, we will solve for the perturbations around the 
$T=0$ asymptotically $AdS_5$ solution also for the remaining functions 
$V_1$, $L_1$, and $M_1$ in order to confirm that the expansion and matching 
procedure works consistently to leading order also for these functions.

\subsection{Solving and matching for $V_1$,  $L_1$, and $M_1$}

The calculation of the perturbations $V_1$ and $L_1$ is considerably more 
complicated than for $N_1$ and $P_1$, because their equations do not decouple. 
In Appendix B, the perturbations of $V$ and of the composite $f=L^2-MN$ 
are obtained from the perturbations of a common field $X$, via the relations,
$e^{2V} = X'' $ and $ f e^{2V} = 24 X$. The zero-th order solution corresponds
to the $L_0$ and $V_0$ fields of the $T=0$ asymptotically $AdS_5$ solution, 
and is given by $24 X_0 = L_0^2 e^{2V_0}$, since $N_0=0$. The perturbations 
$V_1$ and $L_1$ are obtained from the perturbations $X_1$ of $X$,
and the known perturbation $N_1$ by the relations,
\bea
\label{X1V1}
2V_1 & = & X_1 '' e^{-2V_0} 
\no \\
 2 L_0 L_1 & = &  - 2 V_1 L_0^2 + 24 X_1 e^{-2V_0} + M_0 N_1
\eea
It is shown in Appendix B that the field $X_1$ obeys a third order linear differential 
equation, whose inhomogeneous part is sourced by $N_1$ and $P_1$. Two linearly 
independent solutions are known to the homogeneous 
part of this equation, namely the translation and dilation modes of $X_0$, 
which we denote by $X_1^t$ and $X_1^d$ respectively. They are given by,
\bea
X_1 ^t = X_0' \hskip 1in X_1^d = r X_0' - 2 X_0
\eea
The third solution $X_1^n$ to the homogeneous equation, and a particular solution 
$X_1^p$ to the full inhomogeneous $X_1$-equation are constructed in Appendix B, 
so that the general solution for $X_1$, $V_1$ and $L_1$ is given by,
\bea
X_1 & = & \zeta _t X_1^t + \zeta _d X_1^d + \zeta _n X_1^n + \tilde n_0 X_1^p
\no \\
V_1 & = & \zeta _t V_1^t + \zeta _d V_1^d + \zeta _n V_1^n + \tilde n_0 V_1^p
\no \\
L_1 & = & \zeta _t L_1^t + \zeta _d L_1^d + \zeta _n L_1^n + \tilde n_0 L_1^p
\eea
The translation and dilation modes of $V_1$ and $L_1$ (as well as of the 
remaining fields) were already given in (\ref{tdeq}). The $n$ and $p$ modes are
given by
\bea
2V_1^n = (X_1^n) '' e^{-2V_0} 
& \hskip 0.8in &
 2 L_0 L_1^n  =  - 2 V_1^n L_0^2 + 24 X_1^n e^{-2V_0}
 \no \\
2V_1^p = (X_1^p) '' e^{-2V_0} 
&  &
2 L_0 L_1^p  =  - 2 V_1^p L_0^2 + 24 X_1^p e^{-2V_0} + M_0 N_1
\eea
The large $r$ asymptotics of the $X_1$ modes was evaluated in (\ref{X1large}).
The modes $t,n$, and $p$ all respect the asymptotically $AdS_5$ metric,
and are thus allowed from the large $r$ boundary perspective, while the 
mode $d$ violates those boundary conditions and must be absent, so that 
$\zeta _d=0$. As $r \to 0$, the asymptotic behavior of the remaining modes
$X_1^t,X_1^n,X_1^p$ was evaluated in (\ref{X1as}), and we derive the 
following asymptotics for the corresponding $V_1$ and $L_1$ components, 
\bea
V_1^t \sim r^{\sigma -1} ~~ & \hskip 1in & L_1^t \sim r^\sigma
\no \\
V_1^n \sim r^{-\sigma -1} && L_1 ^n \sim r^{-\sigma}
\no \\
V_1^p \sim r^\sigma ~~~~ \, && L_1^p \sim r^{\sigma +1}
\eea
The $V_1^n$ behavior as $ r \to 0$ requires it to match on to the 
$r^{-1-\sigma}$ branch of the solution (\ref{AVV1}) in the overlap region.
As was shown in (\ref{AVV2}), however, the coefficient $v_-$ of this branch
is suppressed by a factor $(mn)^{1+2\sigma} \sim (mn)^{2.5}$ compared to the 
$v_+$ branch and thus corresponds to a higher order effect, which may be neglected.
Thus, the solution in the overlap region instructs us that the $V_1^n$ mode must 
be absent to leading order as well, so that $\zeta _n=0$. It remains to match 
the mode $V_1^p$. Combining it with the leading contribution from the zero-th
order asymptotics $r^\sigma$, we have
\bea
v_+ = (v_0 + nq^2 A_q) V_\sigma (mn)^{-\sigma} = 1 + 2 \tilde n_0 \sigma J_2(k)
\eea
where the coefficient $J_2(k)$ is defined in (\ref{jay2}) of Appendix B.
Clearly, the second term on the rhs is subdominant, and may be dropped in 
the matching process, yielding the matching condition of (\ref{match7}).

\subsubsection{Solving and matching for $M_1$}

The homogeneous part of the only remaining equation E2 is solved by $L_0$ 
and $L_0^c$. The general solution is given by,
\bea
\label{M1int}
M_1(r) & = & \tilde m_0 L_0^c(r) + \tilde m_1 L_0(r)
\no \\ &&
- L_0^c(r)  \int ^r _0 dr'  L_0 e^{2V_0} \left ( 2 V_1' M_0'  + 4 M_0 
\Big ( V_1'' + 2 V_0' V_1' \Big ) + 8 E_0 E_1  \right )
\no \\ &&
+L_0(r)  \int ^r _\infty  dr'   L_0^c
e^{2V_0} \left ( 2 V_1' M_0'  + 4 M_0 
\Big ( V_1'' + 2 V_0' V_1' \Big ) + 8 E_0 E_1  \right )
\eea
The lower boundaries of the integrations are of course arbitrary, since 
their change can always be compensated by changing the integration
constants $\tilde m_0, \tilde m_1$. 

\sm

As $ r \to \infty$, the requirement that the perturbation $V_1$ should maintain 
the asymptotics  $e^{2V} \sim 2c_V r$, requires that $V_1$ must vanish at least
as fast as $1/r$. (The mode $V_1^t, V^n_1$, and $V_1^p$ all satisfy this 
requirement, while the dilation mode does not; thus the dilation mode
must be absent, and we have $\zeta _d=0$.) Given that $M_0$ and $L_0$ 
are linear in $r$ as $ r \to \infty$, while $L_0^c$ goes like $1/r$, 
it follows that the leading asymptotics as $ r \to \infty$ of the integrands is given by,
\bea
L_0 e^{2V_0} \Big ( 2 V_1' M_0'  + 4 M_0  \Big ( V_1'' + 2 V_0' V_1' \Big ) + 8 E_0 E_1 \Big )
& \sim & \cO(r^0)
\no \\
L_0^c e^{2V_0} \Big ( 2 V_1' M_0'  + 4 M_0  \Big ( V_1'' + 2 V_0' V_1' \Big ) + 8 E_0 E_1 \Big )
& \sim & \cO(r^{-2})
\eea
Hence, the second integral is convergent; the first integral behaves linearly
as $ r \to \infty$ so that the contributions of both integral terms behave as
constants when $r \to \infty$. Demanding that $c_M$ be unchanged by the 
perturbation $\ep_T$ then requires that $\tilde m_1=0$.

\subsection{Calculation of $C_T$ at the critical point}

The coefficient $C_T$ was already computed in section \ref{scaling}
by indirect matching methods. For completeness, we shall here also give
its derivation directly from the perturbation theory solutions.
The coefficient $C_T$ is defined by the following relations as $r \to 0$,
\bea
\label{epTdef}
M(r) & \sim & - \tilde \a r - \ep _T r 
\no \\
N & = & C_T \ep _T 
\eea
while keeping $c_E$ constant. Actually, at the critical solution, we have $\tilde \a =0$. 
To determine the coefficient $C_T$, we need both the functions $M_1$ and $E_1$,
the latter because we need to make sure that when turning on $N$ at the horizon, we
keep $c_E$ unchanged. In fact, from formula (\ref{E1as}), it is immediately 
obvious that to keep $c_E$ constant, we need to require
\bea
a_1 =  - n_0 { C_B \over c_{E0}}
\eea
so that the field $E_1=A_1'$ relevant to the $\ep_T$ perturbation at constant $c_E$
is given by
\bea
\label{A1eq}
A_1 (r) = A_0(r) \left [-n_0 {C_B \over c_{E0}}  -  \int _\infty ^r dr' 
{1 \over A_0(r')}  \left ( 2  E_0 V_1 
+ {E_0 \over L_0} L_1 + {M_0 \over L_0} P_1 \right )(r') \right ]
\eea
We begin by obtaining the $r \to 0$ asymptotics of $E_1$. Since we have 
$E_0 (r) / A_0(r) \sim k /r$, the last two terms dominate and their integral 
produces the following leading asymptotics, 
\bea
A_1(r) \sim {n_0 \tilde \a k \over 24 \sqrt{12} (2k-1)} {A_0(r)\over r} + \cO(r^0) A_0(r)
\eea
Now, the key ingredient we use here is that at the critical point, $\tilde \a=0$,
so this singular leading asymptotics actually drops out. (Away from the critical
point, we would have to compensate for this mode by an explicit subtraction 
of a translation mode.) As a result, the leading asymptotics of $E_1$ at the 
critical point is just $E_1(r) \sim r^{k-1}$.
We are now in a position to evaluate the asymptotics of the integrands in 
(\ref{M1int}) as $ r \to 0$, taking $V_1=V_1^p \sim V_0\sim r^\sigma$, 
and $E_1 \sim E_0 \sim r^{k-1}$. The terms in $M_0$ and $M_0'$ scale as $r^\sigma$, 
while the term in $E_1$ scales as $r^{2k-1}$.
As a result, the first integral in (\ref{M1int}) is convergent as $ r \to 0$, but in fact for 
$k>1/2$, it vanishes faster than $r$ as $ r \to 0$, and thus never contributes to the
linear behavior of $M(r)$ which we need to extract following (\ref{epTdef}).
Thus, the first integral is irrelevant to the calculation of $C_T$.  The second integral
in (\ref{M1int}) is convergent as $ r \to 0$ as long as $k>1/2$. Thus, we extract, 
\bea
 {1 \over C_T}  & = & {\sqrt{12} \over n_0} \int ^\infty _0  dr'   L_0^c
e^{2V_0} \left ( 2 V_1' M_0'  + 4 M_0 
\Big ( V_1'' + 2 V_0' V_1' \Big ) + 8 E_0 E_1  \right )
\eea
where $V_1=V_1^p$, and $E_1=A_1'$ is the field defined in (\ref{A1eq}).
All dependence on $n_0$ drops out of these formulas, and the remainder
is expressed solely in terms of the functions of the $T=0$ critical solution 
times a factor of $c_{E0}^4$.

\section{Discussion}
\setcounter{equation}{0}
\label{discussion}

In this paper we have restricted attention to the gravity side of the holographic duality, but
it will of course be illuminating to further explore the field theory interpretation of the 
quantum critical point and its properties.   In \cite{D'Hoker:2010rz} we noted that the 
scaling properties of the entropy density agree with those near a quantum  critical point described by the
Hertz-Millis theory  \cite{hertz,millis93,Millis02}.  To  develop this connection further  it will be very useful
to study finite frequency/wavelength perturbations around the critical point and to compute the
associated dispersion relations.  Based on our experience with finite temperature perturbations, 
it seems likely that this is analytically tractable. 

\sm

Even at the level of free field theory, there are possible origins for the emergence of a
critical magnetic field.   In the simplest interpretation, we assume that for $\Bh> \Bh_c$ the charge
density in the field theory is due to fermions occupying the lowest Landau level.  This is energetically
favorable, as both higher fermionic Landau levels as well as bosonic levels have higher energy. 
As the magnetic field is decreased, or  the charge density is increased, the maximal $p_3$ 
(the momentum parallel to the magnetic field vector) carried
by states in the lowest Landau level increases, which increases their energy, $E = |p_3|$. 
Eventually, the energy exceeds that of the higher fermionic levels and the lowest bosonic level.
By dimensional analysis, this occurs when $B^3/\rho^2$ is of order unity, with the precise number
depending on the charge spectrum.   It is possible that the critical point we see on the gravity
side is in some way connected to this, although  free field theory clearly seems incapable of explaining
the details, such as the $\sh \sim \Th^{1/3}$ scaling law.  

\sm

Another interesting question concerns the effective spacetime dimensionality of the fixed point 
theory, by which we mean the number of dimensions along which long range correlations are
present.   Several features seem to point to a $1+1$ dimensional interpretation, especially the
appearance of near horizon geometries that factorize as $M_3 \times R^2$.  However,  examining
more closely our analysis of the low temperature thermodynamics, we see that this is not so 
clear cut.  At nonzero charge density and magnetic field, and for vanishing $P_3$ chemical 
potential corresponding to the vanishing of $L$ at the horizon,  there is no finite temperature 
solution taking the form of a three dimensional black hole times $R^2$. 
Indeed, in Appendix \ref{nhsolutions} we write down the general factorized solution, and none obeys these
conditions, except when $k= 1$.  Our low temperature solutions always involve some excitation
of $V$, corresponding to a fully five-dimensional solution.   This seems to suggest that the degrees
of freedom contributing to the low temperature entropy density  live in more than $1+1$ dimensions. To
better understand this, it will again be very useful to have results for the spectrum of finite
frequency/wavelength perturbations around the critical point.

\bigskip\bigskip

\noindent
{\Large \bf Acknowledgments}

\medskip

It is a pleasure to thank Eric Perlmutter for discussions.

\appendix 

\section{General solution with constant $V$}
\setcounter{equation}{0}
\label{nhsolutions}

In this appendix we derive the most general solution for which $V$ is constant,
and $b \not=0$. By rescaling $x_1$ and $x_2$ we set $V=0$. 
The reduced field equations are then 
\bea
M1 & \hskip 0.5in &   (NE+LP)' + 2 k b P =0
\no \\
M2 &  &  (LE+MP)' - 2 k b E =0
\no \\
E1 & \hskip 0.5in  &  L'' =  4 EP 
\no \\
E2 &  & M'' =- 4 E^2 
\no \\
E3 &  & N'' =- 4 P^2 
\no \\
E4 &  &  (L')^2 =M'N' + 4 b^2
\no \\
CON &  &  MP^2 +2LEP+NE^2 -6 +{1\over 4}(L')^2 - {1 \over 4} M'N' + b^2 =0
\eea
We begin by eliminating the bilinears in $E$ and $P$ between equations $E1, E2, E3$, 
\bea
\label{A3}
 (L'')^2 - M''N'' =0
\eea
If $N' \not=0$,  we  solve for $M'$ in terms of $L'$ and $N'$ using E4,
\bea
M' = { (L')^2 - 4b^2 \over N'}
\eea
Differentiating once and substituting $M''$ into (\ref{A3}), we obtain 
a factorized equation, 
\bea
\Big ( N'L'' -L'N''-2bN'' \Big ) \Big ( N'L'' -L'N''+2bN'' \Big ) =0
\eea
At least one of these factors must vanish. Without loss of generality, we
choose this to be the first factor (the case of the second factor 
vanishing corresponding to $ b \to -b$), so that $N'L'' - (L'+2b)N'' =0$.
This means that the functions $N' $ and $L' +2b$ must be proportional 
to one another. Since we have assumed that $N'\not=0$, we have in all generality, 
\bea
L' + 2 b = a N'
\eea
for some constant $a$. As a result, we have $L'' = a N''$ which implies
$P(E+aP)=0$ by equations E1 and E3. By an SL(2,R) transformation, 
any solution $E+aP=0$ is equivalent to a solution $P=0$. If $N'=0$, then
we must also have $P=0$. We conclude that all solutions with $V$ constant 
and $b\not=0$ are SL(2,R)-equivalent to a solution with $P=0$. We shall
now proceed to solve the general case where $V$ is constant and $P=0$.

\subsection{The general solutions for $P=0$}

Constant $V$ and $P=0$ simplify E1, E3 and CON to the following relations,
\bea
L''=N''=0 \hskip 1in NE^2=6-2b^2
\eea
the first two of which are solved by
\bea
L(r)&=& l_0 + l_1 r
\no \\
N(r)&=& n_0 + n_1 r
\eea
for constant $l_0, l_1, n_0, n_1$. We now have the following three classes of solutions.
\begin{enumerate}
\item When $n_1=0$, and $ n_0=0$, we have $b^2=3$, with $l_1=\pm 2 b $, and 
the only non-trivial remaining equations are M2 and E2, which are solved by,
\bea
E(r) & = & q r^{\pm k-1}
\no \\
M(r) & = & m_0  +m_1 r  - { 2 q^2 \over k(2k\mp 1)} r^{\pm 2k}
\eea
for arbitrary constants $q$, $m_0,m_1$. This solution  is equivalent to (\ref{nhsol}).
\item When $n_1=0$, and $ n_0 \not=0$, we rescale to $n_0=1$, so that 
$E=q$ is a constant obeying  $q^2 + 2b^2 =6$, with $L=\pm 2b$ from E4,  the 
constraint $q(k\mp 1)=0$ from M2, and
\bea
M(r) = m_0+m_1r - 2 q^2 r^2
\eea
for arbitrary constants $m_0$ and $m_1$. This is equivalent to the $k=\pm 1$
solution of (\ref{kone}).
\item When $n_1 \not=0$, we have $M'$ constant by E4, $E=0$ by E2,  $b^2=3$
by E4 and CON, and
\bea
M(r) = m_0 + m_1 r
\eea
with $l_1^2 -m_1n_1 = 12$. This solution is SL(2,R)-equivalent to $AdS_3$, with a
standard presentation given by $l_1=\sqrt{12}$ and $m_1=n_1=0$.

\end{enumerate}

\section{Solving for $V$ perturbations on asymptotically $AdS_5$}
\setcounter{equation}{0}
\label{Vsection}

To solve for the perturbations $V_1$, given that the perturbations 
$N_1$ and $P_1$ are now known, we need to solve a coupled set of
equations for $V_1$ and $f_1$, such as fV and E4'. Eliminating $f_1$
would result in a 4-th order differential equation in $V_1$,
whose solvability is unclear. Here, instead, we shall take a less direct approach;
it will result in a single third order differential equation for the full composite 
$f \, e^{2V}$; this equation will then be linearized around the $f_0,V_0$ solution;
two of its solutions will be known as the corresponding translation and dilation 
modes, and the third solution may then be constructed by quadrature only. 
 
\sm

Our starting point will be the following two independent equations for the full fields,
\bea
\label{fV5}
0 & = & \left ( f\,  e^{2V} \right )'' - 24 e^{2V}
\\
0 & = & f (V')^2 + f' V' +{g \over 4}  + 3 e^{-4V} - 6
+ MP^2 + 2 LEP + NE^2
\no
\eea
where $f,g$ were defined in (\ref{fg}). The first equation coincides with (\ref{fV}), 
while the second is the original CON equation of (\ref{baneqs}). These two equations 
combined are equivalent to E1, E4, and CON.

\subsection{Derivation of a single third order differential equation}

Equation (\ref{radial}) allows us to express $g$ in terms of $f, f'$ and the combination
$\lambda ^2 - \mu \nu$, which in turn may be evaluated with the help of the first
integrals of (\ref{einstint}). Carrying out these substitutions, we may re-express
the second equation of (\ref{fV5}) as follows, 
\bea
\label{newcon}
&& f (V')^2 + f' V' +{(f')^2  \over 16f } + b^2 e^{-4V} -6
+ MP^2 + 2 LEP + NE^2
\no \\ && \qquad =
  { 1 \over 4f} \left [ \l_0^2 - \mu _0 \nu_0 
+ 4 kb \left ( 2 \l_0 AC + \mu_0 C^2 + \nu _0 A^2 \right ) \right ]\, e^{-4V}
\eea
Next, we solve the first equation in (\ref{fV5}) by parametrizing the most general
solution with the help of a function $X$, such that,
\bea
f \, e^{2V} & = & 24 X
\no \\
e^{2V} & = & X''
\eea
and eliminate $f$ and $V$ in favor of $X$ in (\ref{newcon}). This gives the following
equation,  
\bea
\label{RX}
-3 (XX''')^2 + 6 XX'X''X''' + (X'X'')^2 + 2 XX''-4 X(X'')^3  =R
\eea
where $R$ is defined by
\bea
R& = & 
-  {1 \over 36} (MP^2 + 2 LEP + NE^2)f e^{8V}
\no \\ &&
 + { 1 \over 144} \left [ \l_0^2 - \mu _0 \nu_0 
+ 4 kb \left ( 2 \l_0 AC + \mu_0 C^2 + \nu _0 A^2 \right ) \right ]e^{4V}
\eea
In the perturbative expansion of this equation, $R$ will act as source terms. 
For this reason, we have refrained from recasting its $f$ and $V$ dependence 
in terms of $X$. 

\sm

The unperturbed solution in terms of $L_0,M_0,V_0,A_0$ has $N_0=C_0=P_0=0$.
Its boundary conditions $A_0(0)=L_0(0)=M(0)=0$ force the values of the first 
integrals $\lambda_0=\mu_0=\nu_0=0$, so that $R_0=0$, and $X_0$ obeys
\bea
\label{X0eq}
-3 (X_0X_0''')^2 + 6 X_0X'_0X_0''X_0''' + (X_0'X_0'')^2 + 2 X_0X_0''-4 X_0(X_0'')^3 =0
\eea
First order perturbation theory around this solution was organized 
in (\ref{expeps}), and we set
\bea
X & = & X_0 + \ep X_1
\no \\
R & = & R_0 + \ep \tilde n_0 R_1
\eea
By inspection of (\ref{einstint}), the constants $\lambda _0, \mu_0$ and $\nu_0$
are all at least of order $\ep$ (since they vanished for the unperturbed solution).
Thus,  $R_1$ may be readily evaluated and we find,
\bea
R_1 = -  {1 \over 36 \tilde n_0 } ( 2 L_0E_0P_1 + N_1E_0^2)f_0 e^{8V_0}
 + { kb \nu_0 \over 36 \tilde n_0} A^2_0 e^{4V_0}
\eea
The constant $\nu_0$ is readily deduced using the last equation of (\ref{einstint}),
the expression for $N_1$ in (\ref{N1eq}), as well as (\ref{Lceq}), and we find,
\bea
\nu_0 = - \tilde n_0
\eea
Finally, we linearize also the left side of (\ref{RX}), and organize the result as follows,
\bea
\label{X1eq}
a_3 X_1''' + a_2 X_1'' + a_1 X_1' + a_0 X_1 = \tilde n_0 R_1
\eea
where the coefficients are found to be,
\bea
\label{ass}
a_3 & = & - 6 X_0^2 X_0''' + 6 X_0 X_0' X_0''
\no \\
a_2 & = & 6 X_0 X_0' X_0''' + 2 (X_0')^2 X_0'' + 2 X_0 - 12 X_0 (X_0'')^2
\no \\
a_1 & = & 6 X_0 X_0'' X_0''' + 2 X_0' (X_0'')^2
\no \\
a_0 & = & - 6 X_0 (X_0''')^2 + 6 X_0' X_0'' X_0''' + 2 X_0'' - 4 (X_0'')^3
\eea
It is this equation for $X_1$ than remains to be solved.

\subsection{Constructing the general solution for $X_1$}

Equation (\ref{X0eq}) for the unperturbed solution is invariant under translations 
and dilations in $r$. As a result, the homogeneous 
part of (\ref{X1eq}) manifestly admits two solutions: the collective coordinate 
modes $X_1^t$ and $X_1^d$ associated respectively with translations and dilations
of the unperturbed solution $X_0$. We know their form explicitly in terms of $X_0$,
\bea
X_1 ^t = X_0'
\hskip 1in 
X_1 ^d = r X_0' - 2 X_0
\eea
The fact that $X_1^t$ solves the homogeneous part of (\ref{X1eq}) allows one  
to reduce the order of the differential equation by one. To do so, we introduce
a function $Y$ which satisfies,
\bea
X_1 = X_0' Y
\hskip 1in 
b_3 Y''' + b_2 Y'' + b_1 Y' = \tilde n_0 R_1
\eea
where 
\bea
b_3 & = & a_3 X_0' 
\no \\
b_2 & = & a_2 X_0' + 3 a_3 X_0''
\no \\
b_1 & = & a_1 X_0' + 2 a_2 X_0'' + 3 a_3 X_0 '''
\eea
Since the homogeneous part of equation (\ref{X1eq}) admits also a second solution,
namely $X_1^d$, it follows that the homogeneous part of the equation in $Y$
must admit the solution, 
\bea
Y_0 = { r X_0' - 2 X_0 \over X_0'} = r - 2 { X_0 \over X_0'}
\eea
Introducing now a function $Z$ which satisfies,
\bea
\label{fullZ}
Y' = Y_0' Z
\hskip 1in 
Z'' b_3 Y_0' + Z'(2b_3Y_0''+b_2Y_0') = \tilde n_0 R_1
\eea
The homogeneous part of this equation reduces to,
\bea
{ Z_0'' \over Z_0'} + 2 {Y_0''\over Y_0'} + 3 {X_0'' \over X_0'} + { a_2 \over a_3}=0
\eea
The quantity $a_2/a_3$ may be evaluated in terms of the functions $L_0$
and $V_0$, and may be reduced to the following simple expression,
\bea
{a_2 \over a_3} = -  4 V_0' - {L_0' \over L_0}   - { L_0'' \over L_0'}
\eea
Putting all this together allows us to compute,
\bea
Z_0' = { L_0 L_0' e^{4 V_0} \over (X_0')^3 (Y_0')^2}
\eea
Finally, it remains to find a particular solution to the  inhomogeneous equation
for $Z$ of (\ref{fullZ}). To do so, we introduce a function $Q$ which satisfies,
\bea
Z' = \tilde n_0 Z_0' Q_1 
\hskip 1in 
Q_1' = { R_1 \over a_3 X_0' Y_0' Z_0'}
\eea
Clearly, this equation can be solved by quadratures only.

\sm

We are now ready to present the general solution for $X_1$, 
in terms of its three linearly independent homogeneous solutions
$X_1^t$, $X_1^d$, and the third solution which we shall denote $X_1^n$,
as well as a particular solution $X_1^p$ sourced by $R_1$. They are
given explicitly as follows,
\bea
X_1^t(r) & = & X_0'(r) 
\no \\
X_1 ^d (r) & = &  X_0' (r) Y_0(r) 
\no \\
X_1 ^n (r) & = & X_0'(r)  \int _{\infty} ^r dr' Y_0'(r') Z_0 (r') 
\no \\ 
X_1 ^p (r) & = & 
X_0'(r) \int ^r _0 dr' Y_0'(r') \int ^{r'} _{\infty} dr'' Z_0'(r'') Q_1 (r'')
\eea
where we have defined $Q_1$ to vanish at $r=0$,
\bea
Q_1 (r) =  \int ^r _0 dr' \, Q_1'(r')
\eea
The general solution for $X_1$ is then given by,
\bea
X_1 = \zeta _t X_1^t + \zeta _d X_1^d + \zeta _n X_1^n + \tilde n_0 X^p_1 
\eea
The scripts $t,d,n,p$ refer respectively to translations, dilations, non-local, 
and particular solution, and $\zeta _t$, $\zeta _d$, and $\zeta_n$ are integration
constants.  The normalizations are 
such that $Q_1(0)=0$, $Y_0(0)=0$, and $Z_0(\infty)=0$.

\subsection{Asymptotic behavior of  $X_1$ and $V_1$}

The effects on the fields of the translation and dilation modes are well-known,
and have already been spelled out in (\ref{tdeq}). To derive the asymptotics of the 
$X_1^n$ and $X_1^p$ modes, and their corresponding effects on the other fields,
we express the modes in terms of the unperturbed solution and the fields $N_1$ 
and $P_1$. One finds, 
\bea
X_0' = { 1 \over 12} L_0 (L_0'+L_0 V_0') e^{2V_0}
& \hskip 0.5in &
Y_0' =  { 12 - (L_0'+L_0V_0')^2  \over (L_0'+L_0V_0')^2}
\no \\ &&
Z_0' = { (12)^3 L_0' \over L_0^2 \, e^{2V_0}} \, 
{ L_0' + L_0 V_0' \over \left [ 12 - (L_0'+L_0V_0')^2  \right ]^2} \qquad
\eea
as a result, the equation for $Q_1$ becomes,
\bea
Q_1' & = &  { 12-(L_0'+L_0V_0')^2  \over 3 L_0^2 (L_0')^2 \, e^{6V_0}}  \, R_1
\no \\
R_1 & = & -  {1 \over 36 \tilde n_0 } ( 2 L_0E_0P_1 + N_1E_0^2)L_0^2 e^{8V_0}
 - { kb \over 36} A^2_0 e^{4V_0}
\eea
The $r \to \infty$ asymptotics of $X_1$ may be evaluated using the asymptotics 
of $V_0$, $L_0$, and of $N_1,P_1$ from (\ref{N1as}) and
(\ref{P1as}) respectively, and we find,
\bea
\label{X1large}
X_1 ^t \sim {c_V \over 2} r^2 & \hskip 0.6in & X_1^n \sim 24 r
\no \\
X_1^d \sim {c_V \over 6} r^3 & \hskip 0.6in & 
X_1^p \sim  J_1(k) X_1^t +  Q_1(\infty) X_1^n
\eea
where $J_1(k)$ is defined by the integral, 
\bea
J_1(k) \equiv \int ^\infty _0 dr' Y_0'(r') \int ^{r'} _{\infty} dr'' Z_0'(r'') Q_1 (r'')
\eea
which is convergent for all $k >1/2$. Clearly, the boundary conditions 
as $r \to \infty$ require the dilation mode to be absent, so that $\zeta _d=0$,
but allows for the three remaining modes.

\sm

The $r \to 0$ asymptotics of $X_1$ may be evaluated similarly. Actually, we shall be
interested only in contributions to $X_1$ that produce functional dependences in $V_1$
of the form $r^\sigma$ and $r^{-1-\sigma}$ in this limit, since these are the only modes that
enter in the overlap region of (\ref{overlap}). Since we have $2V_1 e^{2V_0} =  X_1''$,
the corresponding modes in $X_1$ behave as $r^{\sigma +2}$ and $r ^{1-\sigma}$.
Using the $r \to 0$ asymptotics of $L_0, V_0$, and $N_1,P_1$ from (\ref{N1as}) and
(\ref{P1as}) respectively, we find,
\bea
\label{X1as}
X_1 ^t \sim r \hskip 0.72in  & \hskip 0.6in & 
X_1^n \sim { 6 \, r^{1-\sigma} \over \sigma (2\sigma +1) (3 \sigma -2)}
\no \\
X_1^d \sim { 6 \sigma -4 \over \sigma +1} r^{2 + \sigma} & \hskip 0.6in & 
X_1^p \sim  J_2(k) X_1^d
\eea
where $J_2(k)$ is defined by the integral
\bea
\label{jay2}
J_2(k) \equiv - \int _0 ^\infty dr Z_0'(r) Q_1(r)
\eea


\newpage




\begin{thebibliography}{99} 


\bibitem{Liu:2009dm}
  H.~Liu, J.~McGreevy and D.~Vegh,
  ``Non-Fermi liquids from holography,''
  arXiv:0903.2477 [hep-th].

\bibitem{Cubrovic:2009ye}
  M.~Cubrovic, J.~Zaanen and K.~Schalm,
  ``String Theory, Quantum Phase Transitions and the Emergent Fermi-Liquid,''
  Science {\bf 325}, 439 (2009)
  [arXiv:0904.1993 [hep-th]].

\bibitem{Faulkner:2009wj}
  T.~Faulkner, H.~Liu, J.~McGreevy and D.~Vegh,
  ``Emergent quantum criticality, Fermi surfaces, and AdS2,''
  arXiv:0907.2694 [hep-th].

\bibitem{Faulkner:2010tq}
  T.~Faulkner and J.~Polchinski,
  ``Semi-Holographic Fermi Liquids,''
  arXiv:1001.5049 [hep-th].

\bibitem{Sakai:2004cn}
  T.~Sakai and S.~Sugimoto,
  ``Low energy hadron physics in holographic QCD,''
  Prog.\ Theor.\ Phys.\  {\bf 113}, 843 (2005)
  [arXiv:hep-th/0412141].

\bibitem{Davis:2008nv}
  J.~L.~Davis, P.~Kraus and A.~Shah,
  ``Gravity Dual of a Quantum Hall Plateau Transition,''
  JHEP {\bf 0811}, 020 (2008)
  [arXiv:0809.1876 [hep-th]].



\bibitem{Rey:2008zz}
  S.~J.~Rey,
  ``String Theory On Thin Semiconductors: Holographic Realization Of Fermi
 Points And Surfaces,''
  Prog.\ Theor.\ Phys.\ Suppl.\  {\bf 177}, 128 (2009)
  [arXiv:0911.5295 [hep-th]].


  
\bibitem{Kulaxizi:2008jx}
  M.~Kulaxizi and A.~Parnachev,
  ``Holographic Responses of Fermion Matter,''
  Nucl.\ Phys.\  B {\bf 815}, 125 (2009)
  [arXiv:0811.2262 [hep-th]].


\bibitem{D'Hoker:2009mm}
  E.~D'Hoker and P.~Kraus,
  ``Magnetic Brane Solutions in AdS,''
  JHEP {\bf 0910}, 088 (2009)
  [arXiv:0908.3875 [hep-th]].

\bibitem{D'Hoker:2009bc}
  E.~D'Hoker and P.~Kraus,
  ``Charged Magnetic Brane Solutions in AdS$_5$ and the fate of the third law of
  thermodynamics,''
  JHEP {\bf 1003}, 095 (2010)
  [arXiv:0911.4518 [hep-th]].

\bibitem{D'Hoker:2010rz}
  E.~D'Hoker and P.~Kraus,
  ``Holographic Metamagnetism, Quantum Criticality, and Crossover Behavior,''
  JHEP {\bf 1005}, 083 (2010)
  [arXiv:1003.1302 [hep-th]].

\bibitem{LRMW}
H. v. Lohneysen, A. Rosch, M. Mojta, and P. Wolfle, ``Fermi-liquid instabilities at magnetic 
quantum phase transitions", 
Rev. Mod. Phys. {\bf 79} , 1015–1075  (2007) 

\bibitem{RPMMG}
A. W. Rost, R. S. Perry, J.-F. Mercure, A. P. Mackenzie, and  S. A. Grigera
``Entropy Landscape of Phase Formation Associated with Quantum Criticality in 
$Sr_3Ru_2O_7$", Science Vol. 325. no. 5946, pp. 1360 - 1363 (2009)


\bibitem{Son:2008ye}
  D.~T.~Son,
  ``Toward an AdS/cold atoms correspondence: a geometric realization of the
  Schroedinger symmetry,''
  Phys.\ Rev.\  D {\bf 78}, 046003 (2008)
  [arXiv:0804.3972 [hep-th]].

\bibitem{Balasubramanian:2008dm}
  K.~Balasubramanian and J.~McGreevy,
  ``Gravity duals for non-relativistic CFTs,''
  Phys.\ Rev.\ Lett.\  {\bf 101}, 061601 (2008)
  [arXiv:0804.4053 [hep-th]].


\bibitem{Anninos:2010pm}
  D.~Anninos, G.~Compere, S.~de Buyl, S.~Detournay and M.~Guica,
  ``The Curious Case of Null Warped Space,''
  arXiv:1005.4072 [hep-th].

\bibitem{Brown:1986nw}
  J.~D.~Brown and M.~Henneaux,
  ``Central Charges in the Canonical Realization of Asymptotic Symmetries: An
  Example from Three-Dimensional Gravity,''
  Commun.\ Math.\ Phys.\  {\bf 104}, 207 (1986).

\bibitem{Lippert}
  G.~Lifschytz and M.~Lippert,
  ``Holographic Magnetic Phase Transition,''
  Phys.\ Rev.\  D {\bf 80}, 066007 (2009)
  [arXiv:0906.3892 [hep-th]].

\bibitem{Evans:2010iy}
  N.~Evans, A.~Gebauer, K.~Y.~Kim and M.~Magou,
  ``Holographic Description of the Phase Diagram of a Chiral Symmetry Breaking
  Gauge Theory,''
  arXiv:1002.1885 [hep-th].


\bibitem{Jensen:2010vd}
  K.~Jensen, A.~Karch and E.~G.~Thompson,
  ``A Holographic Quantum Critical Point at Finite Magnetic Field and Finite
  Density,''
  arXiv:1002.2447 [hep-th].



\bibitem{Kachru:2008yh}
  S.~Kachru, X.~Liu and M.~Mulligan,
  ``Gravity Duals of Lifshitz-like Fixed Points,''
  Phys.\ Rev.\  D {\bf 78}, 106005 (2008)
  [arXiv:0808.1725 [hep-th]].



\bibitem{Horowitz:2009ij}
  G.~T.~Horowitz and M.~M.~Roberts,
  ``Zero Temperature Limit of Holographic Superconductors,''
  JHEP {\bf 0911}, 015 (2009)
  [arXiv:0908.3677 [hep-th]].

\bibitem{inprog}
E.~D'Hoker and P.~Kraus,  work in progess

\bibitem{Bobev:2010de}
  N.~Bobev, A.~Kundu, K.~Pilch and N.~P.~Warner,
  ``Supersymmetric Charged Clouds in AdS$_5$,''
  arXiv:1005.3552 [hep-th].

\bibitem{Henningson:1998gx}
  M.~Henningson and K.~Skenderis,
  ``The holographic Weyl anomaly,''
  JHEP {\bf 9807}, 023 (1998)
  [arXiv:hep-th/9806087].

\bibitem{Balasubramanian:1999re}
  V.~Balasubramanian and P.~Kraus,
  ``A stress tensor for anti-de Sitter gravity,''
  Commun.\ Math.\ Phys.\  {\bf 208}, 413 (1999)
  [arXiv:hep-th/9902121].

\bibitem{Buchel:2006gb}
  A.~Buchel and J.~T.~Liu,
  ``Gauged supergravity from type IIB string theory on Y(p,q) manifolds,''
  Nucl.\ Phys.\  B {\bf 771} (2007) 93
  [arXiv:hep-th/0608002].

\bibitem{Gauntlett:2006ai}
  J.~P.~Gauntlett, E.~O Colgain and O.~Varela,
  ``Properties of some conformal field theories with M-theory duals,''
  JHEP {\bf 0702} (2007) 049
  [arXiv:hep-th/0611219].

\bibitem{Gauntlett:2007ma}
  J.~P.~Gauntlett and O.~Varela,
  ``Consistent Kaluza-Klein Reductions for General Supersymmetric AdS
  Solutions,''
  Phys.\ Rev.\  D {\bf 76} (2007) 126007
  [arXiv:0707.2315 [hep-th]].


\bibitem{Clement:1993kc}
  G.~Clement,
  ``Classical solutions in three-dimensional Einstein-Maxwell cosmological
  gravity,''
  Class.\ Quant.\ Grav.\  {\bf 10}, L49 (1993).


\bibitem{Banados:2005da}
  M.~Banados, G.~Barnich, G.~Compere and A.~Gomberoff,
  ``Three dimensional origin of Goedel spacetimes and black holes,''
  Phys.\ Rev.\  D {\bf 73}, 044006 (2006)
  [arXiv:hep-th/0512105].





\bibitem{Anninos:2008fx}
  D.~Anninos, W.~Li, M.~Padi, W.~Song and A.~Strominger,
  ``Warped AdS$_3$ Black Holes,''
  JHEP {\bf 0903}, 130 (2009)
  [arXiv:0807.3040 [hep-th]].


\bibitem{Compere:2009zj}
  G.~Compere and S.~Detournay,
  ``Boundary conditions for spacelike and timelike warped AdS$_3$ spaces in
  topologically massive gravity,''
  JHEP {\bf 0908}, 092 (2009)
  [arXiv:0906.1243 [hep-th]].



\bibitem{hertz}
J. A. Hertz, ``Quantum critical phenomena," Phys. Rev. {\bf B 14}  1165 (1976).

\bibitem{millis93}
A. J. Millis, ``Effect of a nonzero temperature on quantum critical points in itinerant 
fermion systems",
Phys. Rev. {\bf  B 48}, 7183–7196 (1993) 



\bibitem{Millis02}
A. J. Millis, A. J. Schofield, G. G. Lonzarich and S. A. Grigera, ``Metamagnetic quantum
criticality in metals," Phys. Rev. Lett. {\bf 88}, 217204 (2002).

 
\end{thebibliography}
\end{document}